\documentclass[12pt,letterpaper]{article}
\pdfoutput=1

\usepackage{amsmath,amssymb,calc, amsthm,bbm, epsfig,psfrag, mathtools}
\usepackage{graphicx, enumerate}
\usepackage[numbers,sort&compress]{natbib}
\usepackage[dvipsnames]{xcolor}
\usepackage{tensor}

\usepackage[utf8]{inputenc} 
\allowdisplaybreaks

\newcommand{\Comment}[1]{{}}
\definecolor{darkblue}{rgb}{0.15,0.35,0.55}
\definecolor{reddish}{rgb}{0.65, 0.2, 0.2}
\usepackage[linktocpage=true]{hyperref}
\hypersetup{
colorlinks=true,
citecolor=darkblue,
linkcolor=reddish,
urlcolor=darkblue,
pdfauthor={},
pdftitle={},
pdfsubject={}
}

\makeatletter
\renewcommand\section{\@startsection {section}{1}{\z@}%
                                   {-3.5ex \@plus -1ex \@minus -.2ex}
                                   {2.3ex \@plus.2ex}%
                                   {\normalfont\large\bfseries}}
\renewcommand\subsection{\@startsection{subsection}{2}{\z@}%
                                     {-3.25ex\@plus -1ex \@minus -.2ex}%
                                     {1.5ex \@plus .2ex}%
                                     {\normalfont\bfseries}}
\makeatother

\let\non\nonumber

\newcommand{\bea}{\begin{eqnarray}}
\newcommand{\eea}{\end{eqnarray}}

\newcommand{\be}{\begin{equation}}
\newcommand{\ee}{\end{equation}}

\newcommand{\bma}{\begin{pmatrix}}
\newcommand{\ema}{\end{pmatrix}}

\newcommand{\bsubeq}{\begin{subequations}}
\newcommand{\esubeq}{\end{subequations}}

\newfont{\goth}{ygoth.tfm scaled 1200}                   

 \numberwithin{equation}{section}

\def\1{{(1)}}
\def\2{{(2)}}
\def\3{{(3)}}

\def\s{\sigma}

\newcommand\Tb{\overline{T}}
\newcommand\Lb{\overline{L}}
\newcommand\calT{\mathcal{T}}
\newcommand\calTb{\overline{\mathcal{T}}}
\newcommand\calY{\mathcal{Y}}
\newcommand\calYb{\overline{\mathcal{Y}}}
\newcommand\Sb{\overline{S}}
\newcommand\Wb{\overline{W}}
\newcommand\Fb{\overline{F}}
\newcommand\xb{\overline{x}}
\newcommand\Mb{\overline{M}}

\newcommand\phib{\overline{\phi}}
\newcommand\chib{\overline{\chi}}

\newcommand\alphad{\dot{\alpha}}
\newcommand\Phib{\overline{\Phi}}
\newcommand\bbPhi{\pmb{\Phi}}
\newcommand\bbPhib{\overline{\pmb{\Phi}}}
\newcommand\psib{\overline{\psi}}
\newcommand\Db{\overline{D}}
\newcommand\thetab{\overline{\theta}}

\newcommand\nablab{\overline{\nabla}}

\def\a{\alpha}

\def\d{\delta}

\def\q{\theta}

\def\s{\sigma}

\def\z{\zeta}

\def\F{\Phi}

\newcommand{\pa}{\partial}
\newcommand{\qb}{{\bar{\theta}}}
\newcommand{\hf}{\frac12}
\newcommand {\cY}{{\cal Y}}
\newcommand {\cS}{{\cal S}}
\newcommand {\cN}{{\cal N}}
\newcommand {\cU}{{\cal U}}
\newcommand {\cV}{{\cal V}}
\newcommand {\cVb}{{\overline{\cal V}}}

\newcommand{\cO}{{\mathcal O}}

\newcommand{\cF}{{\mathcal F}}
\newcommand{\cQ}{{\mathcal Q}}
\newcommand{\cJ}{{\mathcal J}}

\begin{document}
\begin{titlepage}

\begin{center}

{May 31, 2019}
\hfill         \phantom{xxx}  EFI-19-5

\vskip 2 cm {\Large \bf \texorpdfstring{$T \Tb$}{T Tbar} Flows and \texorpdfstring{$(2,2)$}{(2,2)} Supersymmetry} 
\vskip 1.25 cm {\bf Chih-Kai Chang$^1$, Christian Ferko$^1$, Savdeep Sethi$^1$, 
Alessandro Sfondrini$^2$ and Gabriele Tartaglino-Mazzucchelli$^{3,4}$ }\non\\
\vskip 0.2 cm
 {\it $^1$ Enrico Fermi Institute \& Kadanoff Center for Theoretical Physics \\ University of Chicago, Chicago, IL 60637, USA}
 
\vskip 0.2 cm
 {\it $^2$ Institut f\"ur theoretische Physik, ETH Z\"urich\\ Wolfgang-Pauli-Strasse 27, 8093 Z\"urich, Switzerland}
 
\vskip 0.2 cm
 {\it $^3$ Albert Einstein Center for Fundamental Physics,
Institute for Theoretical Physics,\\
University of Bern,
Sidlerstrasse 5, CH-3012 Bern, Switzerland}

\vskip 0.2 cm
{\it $^4$ School of Mathematics and 
Physics, University of Queensland\\
St Lucia, Brisbane, Queensland 4072, 
Australia}

\vskip 0.2 cm

\end{center}
\vskip 1.5 cm

\begin{abstract}
\baselineskip=18pt

We construct a solvable deformation of  two-dimensional theories with $(2,2)$ supersymmetry using an irrelevant operator which is a bilinear in the supercurrents. This supercurrent-squared operator is manifestly supersymmetric, and equivalent to $T \Tb$ after using conservation laws. 
As illustrative examples, we deform theories involving a single $(2,2)$ chiral superfield. We show that the deformed free theory is on-shell equivalent to the $(2,2)$ 
Nambu-Goto action. 
At the classical level, models with a superpotential exhibit more surprising behavior: the deformed theory exhibits poles in the physical potential which 
modify the vacuum structure. This suggests that irrelevant deformations of $T \Tb$ type might also affect infrared physics.

\end{abstract}

\end{titlepage}

\tableofcontents

\section{Introduction} \label{sec:intro}

Understanding the space of quantum field theories (QFTs) is a fascinating question. A typical approach to this question is to start with 
a particularly tractable model like a free, conformal, or exactly solvable theory, and deform it infinitesimally by adding an integrated local 
operator.  An infinitesimal \textit{relevant} deformation generates a renormalization group flow. 
The resulting theory will differ in the infrared from the original undeformed theory. When the original theory is conformal, there might exist
exactly \textit{marginal} deformations which preserve the conformal symmetry for finite values of the deformation 
parameters; the space of marginal parameters defines the moduli space of the conformal field theory. Finally, if the deforming operator is \textit{irrelevant}, the 
ultraviolet properties of the theory change, and it is usually difficult to understand this change in terms of any kind of flow. This case is the most difficult to understand because, in essence, the definition of the theory changes. 

Irrelevant deformations of two-dimensional Poincar\'e-invariant QFTs generated by the determinant of the stress-energy tensor, 
$\det(T)=T_{00}T_{11}-T_{01}T_{10}$, are special.  These 
$T \Tb$ deformations define a 
flow along which certain properties of the deformed theory can be computed exactly~\cite{Zamolodchikov:2004ce}. Most 
important is the energy spectrum~\cite{Smirnov:2016lqw,Cavaglia:2016oda}. However, in many cases the classical 
action can also be determined in closed form along the flow~\cite{Cavaglia:2016oda,Bonelli:2018kik}.  
This prompted the study of $T \Tb$ deformations of integrable theories~\cite{Cavaglia:2016oda, Conti:2019dxg,Conti:2018jho}, as 
well as of more general theories~\cite{Chen:2018eqk,Datta:2018thy,Aharony:2018bad,Cardy:2018jho,Araujo:2018rho}, with a number of applications 
to (effective) string theory~\cite{Dubovsky:2012wk,Caselle:2013dra,Chen:2018keo}, two-dimensional 
gravity~\cite{Dubovsky:2017cnj,Dubovsky:2018bmo,Conti:2018tca,Ishii:2019uwk} and to the AdS$_3$/CFT$_2$ correspondence~\cite{McGough:2016lol,Giveon:2017nie,Giveon:2017myj, Asrat:2017tzd, Baggio:2018gct,Dei:2018mfl,Gorbenko:2018oov,Dei:2018jyj, Giveon:2019fgr,Giribet:2017imm}.%
\footnote{More general ``$T\overline{J}$'' deformations, which break Lorentz invariance, have also been 
considered~\cite{Guica:2017lia,Bzowski:2018pcy,Nakayama:2018ujt,Chakraborty:2018vja,LeFloch:2019rut,
Guica:2019vnb,Chakraborty:2019mdf}.}

One of the first examples studied was the deformation of a theory of free bosons, which resulted in the Nambu-Goto 
action~\cite{Cavaglia:2016oda}.%
\footnote{This can also be seen by studying the world-sheet S-matrix of strings in flat space~\cite{Dubovsky:2012wk,Caselle:2013dra}. For a theory of free bosons and fermions, one instead finds the Green-Schwarz action in light-cone 
gauge~\cite{Baggio:2018rpv}. See also refs.~\cite{Baggio:2018gct,Frolov:2019nrr} for a discussion of the relation between light-cone 
gauge-fixed strings and $T \Tb$ deformations.}
Interestingly, the $T \Tb$ deformed action for a scalar theory with an arbitrary potential can also be exactly constructed, at least classically. 
Imposing the $T \Tb$ flow equation for the Lagrangian
\begin{equation}
\label{eq:floweq}
    \frac{d}{d\lambda}\mathcal{L}_{\lambda} = \det(T[\mathcal{L}_{\lambda}])\,,
\end{equation}
where the stress-energy tensor $T_{\mu\nu}[\mathcal{L}_{\lambda}]$ is computed in the deformed theory itself, and setting the initial 
condition
\begin{equation}
    \mathcal{L}_{0}= \frac{1}{2}\partial_{++}\phi\partial_{--}\phi + V(\phi)\,,
\end{equation}
gives~\cite{Cavaglia:2016oda,Bonelli:2018kik}:
\begin{equation}
    \mathcal{L}_{\lambda}= \frac{1}{2\lambda}\frac{1-2\lambda V(\phi)}{1-\lambda V(\phi)}\Bigg[-1+
    \sqrt{1+2\lambda \frac{\big(\partial_{++}\phi\partial_{--}\phi + 2V(\phi)\big)
    \big(1-\lambda V(\phi)\big)}{\big(1-2\lambda V(\phi)\big)^2}}\Bigg]
    \,.
\end{equation}
This Lagrangian is fairly involved. It is interesting to consider the potential energy at zero momentum, which means 
discarding all interaction terms which involve derivatives. This captures the potential for slowly-varying fields, 
\begin{equation}
\label{eq:potential-pole}
    \mathcal{L}_{\lambda}= \frac{1}{2}\partial_{++}\phi\partial_{--}\phi+\frac{V(\phi)}{1-\lambda V(\phi)} + \ldots
    \,,
\end{equation}
where the ellipsis denotes interaction terms involving derivatives. Although this is just a classical result, the form of the deformed 
potential is striking: if we start from a regular potential, we will generically develop poles for sufficiently large~$|\lambda|$. These poles 
are invisible in perturbation theory in the flow parameter $\lambda$. 
Were we able to trust this result at the quantum level, this would point to a dramatic modification of the 
theory. Namely, an \textit{irrelevant} deformation would end up changing the \textit{infrared} structure of the 
theory, resulting in a kind of ``IR/UV mixing.''

It is generally not possible to draw firm conclusions about the quantum properties of a theory by studying its classical potential. A truly quantum analysis would certainly be preferable. Unfortunately, our current understanding of the $T \Tb$ deformation at the quantum level is far from complete. When the theory is studied in infinite volume, the deformation can be defined by postulating that the S-matrix only change by a Castillejo-Dalitz-Dyson (CDD) factor~\cite{Cavaglia:2016oda}. This CDD factor, however, spoils the analytic properties of the scattering matrix at large values of the deformation parameter, taking us away from the framework of local QFT. In finite volume, on the other hand, 
a flow equation for the energy spectrum follows from \eqref{eq:floweq}~\cite{Smirnov:2016lqw}.%
\footnote{%
For integrable theories, such a flow equation may also be derived from the CDD deformation using the thermodynamic Bethe ansatz~\cite{Cavaglia:2016oda}.
}
However, generically along the flow some energy levels will become complex; this phenomenon is not completely understood. Which physical observables make sense in the deformed theory is also currently rather mysterious. All in all, a rigorous exploration of possible IR/UV mixing requires a deeper understanding of the quantum properties of the deformed theory.

One instance where a \textit{classical} analysis of the potential might allow us to draw more reliable conclusions about 
the quantum theory is for models with extended supersymmetry. As long as there is sufficient supersymmetry for the 
potential to be partly controlled by a holomorphic quantity, there will be partial protection from perturbative (and sometimes non-perturbative) 
quantum effects. 
Models with $\mathcal{N}=(2,2)$ supersymmetry in two dimensions are precisely of this type, provided that the $T \Tb$ 
deformation is compatible with manifest   $\mathcal{N}=(2,2)$ supersymmetry. 

Recently it was shown that the $T\Tb$ flow preserves manifest
$\mathcal{N}=(0,1)$, $\mathcal{N}=(1,1)$~\cite{Baggio:2018rpv,Chang:2018dge} and $\mathcal{N}=(0,2)$
supersymmetry~\cite{Jiang:2019hux}. Specifically, we can view the flow as generated by the supersymmetric descendant of a composite operator; this composite operator is built from a bilinear in supercurrents.  
This construction both ensures supersymmetry along the flow, and is sufficient to reproduce, and indeed slightly generalize,  Zamolodchikov's argument for the well-definedness and solvability of $T\Tb$~\cite{Zamolodchikov:2004ce}. 
Moreover, for some simple supersymmetric actions it was possible to explicitly construct the deformed Lagrangian in 
superspace, gaining some insight on the resulting theory~\cite{Baggio:2018rpv,Chang:2018dge,Jiang:2019hux}.

The main aim of this paper is to repeat this analysis in the $\mathcal{N}=(2,2)$ case, and find a manifestly $\mathcal{N}=(2,2)$ supersymmetric version of the $T\Tb$ flow. 
The case of $\mathcal{N}=(2,2)$ is particularly interesting for at least two reasons: first, it is the most heavily studied class of two-dimensional supersymmetric theories because of applications to string compactifications. Secondly, these models are closely connected to the dimensional reduction of $\mathcal{N}=1$ theories in four dimensions. Understanding more about the structure of the $\mathcal{N}=(2,2)$ theory might shed light on how to generalize $T\Tb$ to higher dimensions; see~\cite{Taylor:2018xcy,Hartman:2018tkw, Chang:2018dge, Caputa:2019pam} for discussions of such higher-dimensional generalizations. We plan to report on results along this direction in~\cite{SUSYDBITTbar}.

In this work, we will establish the appearance of a singularity in the physical potential, like the one appearing in eq.~\eqref{eq:potential-pole}, 
in a manifestly $\mathcal{N}=(2,2)$ form---where, as usual, the role of $V(\phi)$ will be played by $|W'(\phi)|^2$ with $W(\phi)$ the holomorphic superpotential.
 Our results on the $\mathcal{N}=(2,2)$ version of $T \Tb$ provide a stepping stone toward a fully quantum analysis of the vacuum structure of non-conformal $T \Tb$-deformed theories, which we plan to explore in the future.

The paper is structured as follows: in section~\ref{sec:supercurrents} we review the structure of the $\mathcal{N}=(2,2)$ 
supercurrent multiplets which we need to construct the supersymmetric deformation. In section~\ref{secTTbar} we construct the 
supercurrent-squared operator $\mathcal{T}\overline{\mathcal{T}}$
as a bilinear in the supercurrents and discuss its well-definedness. Finally, in 
section~\ref{sec:models} we construct the deformed action for a few examples of $\mathcal{N}=(2,2)$ theories. In particular, we focus
 on theories involving a single chiral multiplet with an action determined by an arbitrary K\"ahler potential, as well as models with a superpotential. In Appendices~\ref{components-S}, \ref{appendix:supercurrent_calculation} and~\ref{appendix:on-shell}, we collect assorted results used in the main body of the text.
 
\section{\texorpdfstring{$D=2 \;\, \cN=(2,2)$}{2D N=(2,2)} Supercurrent Multiplets} \label{sec:supercurrents}

Our manifestly supersymmetric modification of $T \Tb$ is built from bilinears in fields of the supercurrent multiplet. 
In this section we review the structure of such multiplets in $D=2$ $\mathcal{N} = (2,2)$ theories.

\subsection{Conventions}

We work in two-dimensional $\mathcal{N} = (2,2)$ superspace with Lorentzian signature, see \cite{Gates:1984nk} 
for a classic reference on the subject. 
Our four anti-commuting coordinates are  written $\theta^{\pm}$ and $\thetab^{\pm}$, and we will collectively denote the superspace 
coordinates by $\z^M=(x^\mu,\,\q^\pm,\,\qb^\pm)$.  To more easily interpret expressions involving both vector and spinor quantities, 
we  change to light-cone coordinates using the bi-spinor conventions
\begin{align}
    x^{\pm \pm} = \frac{1}{\sqrt{2}} \left( x^0 \pm x^1 \right) , 
\end{align}
and define the corresponding partial derivatives
\begin{align}
    \partial_{\pm \pm} &= \frac{1}{\sqrt{2}} \left( \partial_0 \pm \partial_1 \right) , 
\end{align}
so that $\partial_{\pm \pm} x^{\pm \pm} = 1$ and $\partial_{\pm \pm} x^{\mp \mp} = 0$. 

Spinors in two dimensions carry a single index which is raised or lowered as follows:
\begin{align}
    \psi^+ = - \psi_-, \qquad   \psi^- = \psi_+ . 
\end{align}
We write all vector indices as pairs of spinor indices. This allows us to nicely compare terms in equations involving combinations of 
spinor, vector, spinor-vector, and tensor quantities. Using this notation, for example, the supercurrent has components 
$S_{+++}, S_{---}, S_{+--}$, and $S_{-++}$, which we can immediately identify as a spinor-vector because it has three indices. 
Similarly, the stress-energy tensor carries two vector indices which are repackaged into four bispinor indices 
$T_{++++}, T_{----}, T_{++--} = T_{--++}$.

The supercovariant derivatives, collectively denoted by $D_A=(\pa_a,\,D_\pm,\Db_\pm)$, are defined by
\begin{align}
    D_{\pm} = \frac{\partial}{\partial \theta^{\pm}} - \frac{i}{2} \thetab^{\pm} \partial_{\pm \pm} 
    ~, ~~~~~~
    \Db_{\pm} = - \frac{\partial}{\partial \thetab^{\pm}} + \frac{i}{2} \theta^{\pm} \partial_{\pm \pm} ~, 
\end{align}
and satisfy
\begin{align}
    \{ D_{\pm} , \Db_{\pm} \} &= i \partial_{\pm \pm} , 
\end{align}
with all other (anti-)commutators vanishing.

The supersymmetry transformations for an $\cN=(2,2)$ superfield  $\cF({\z})=\cF(x^{\pm\pm},\q^\pm,\qb^\pm)$  are given by
\bea
\d_Q \cF
:=
i\epsilon^+ \cQ_+ \cF
+i\epsilon^- \cQ_- \cF
-i\bar\epsilon^+ \overline{\cQ}_+ \cF
-i\bar\epsilon^- \overline{\cQ}_- \cF
~,
\label{susySuperfield22}
\eea
where on superfields the supercharges are represented by the following differential operators
\begin{align}
    \cQ_{\pm} = \frac{\partial}{\partial \theta^{\pm}} + \frac{i}{2} \thetab^{\pm} \partial_{\pm \pm} 
    ~, ~~~~~~
    \overline{\cQ}_{\pm} = - \frac{\partial}{\partial \thetab^{\pm}} - \frac{i}{2} \theta^{\pm} \partial_{\pm \pm} ~, 
\end{align}
satisfying
\begin{align}
    \{ \cQ_{\pm} , \overline{\cQ}_{\pm} \} &= -i \partial_{\pm \pm} 
    ~, 
\end{align}
and commuting with the covariant derivatives $D_A$.

\subsection{The \texorpdfstring{$\cS$}{S}-multiplet}
\label{section-S-multiplet}

For Lorentz invariant supersymmetric theories, there is an essentially unique 
supermultiplet which contains the stress-energy tensor $T_{\mu \nu}$, the supercurrent $S_{\mu \alpha}$, and no other operators 
with spin larger than one, under the assumption that the multiplet, though in general reducible, cannot be separated into decoupled supersymmetry multiplets; 
namely that it is 
indecomposable~\cite{Dumitrescu:2011iu}. This $\cS$-multiplet can be defined in any theory with $D=2$ 
$\cN=(2,2)$ supersymmetry. By ``essentially unique,'' we mean that the $\cS$-multiplet 
is unique up to improvement terms which preserve the superspace constraint equations.

For two-dimensional theories with $(2,2)$ supersymmetry, the $\cS$-multiplet consists of superfields 
$\mathcal{S}_{\pm \pm}$, $\chi_{\pm}$, and $\mathcal{Y}_{\pm}$ which satisfy the constraints:
\bsubeq
\label{conservation-S}
\begin{gather}
    \Db_{\pm} \mathcal{S}_{\mp \mp} 
    = \pm \left( \chi_{\mp} + \mathcal{Y}_{\mp} \right)~ , \\
    \Db_{\pm} \chi_{\pm} = 0~ , 
    \qquad 
    \Db_{\pm} \chi_{\mp} = \pm C^{(\pm)} ~, 
    \qquad 
    D_+ \chi_- - \Db_- \chib_+ = k ~, 
    \\
    D_{\pm} \mathcal{Y}_{\pm} = 0~ ,
    \qquad 
    \Db_{\pm} \mathcal{Y}_{\mp} = \mp C^{(\pm)}~, 
    \qquad 
    D_+ \mathcal{Y}_- + D_- \mathcal{Y}_+ = k' ~.
\end{gather}
\esubeq
Here $k$ and $k'$ are real constants and $C^{(\pm)}$ is a complex constant.
The~$\cS$-multiplet contains~$8+8$ independent real component operators and the constants~$k, k', C^{(\pm)}$ 
\cite{Dumitrescu:2011iu}.
The expansion in components of $\cS_{\pm\pm}$,
$\chi_\pm$, and $\calY_{\pm}$ are given 
for convenience in Appendix
\ref{components-S}.

Among the various component fields 
it is important to single out 
the complex supersymmetry current $S_{\a}{}_\mu$
 and the energy-momentum tensor $T_{\mu\nu}$.
 The complex supersymmetry current, associated to $S_{+\pm\pm}$ and $S_{-\pm\pm}$, is conserved: $\pa^\mu S_{\a}{}_\mu=0$. 
 The energy-momentum tensor,
 associated with $T_{\pm\pm\pm\pm}$ and $T_{++--}=T_{--++}$, is real, conserved ($\pa^\mu T_{\mu\nu}=0$), 
 and symmetric ($T_{\mu\nu}=T_{\nu\mu}$). 
 In light-cone notation the conservation equations are given by
\bsubeq
\label{conserT}
\bea
 \pa_{++} S_{+--}(x) &=&\, - \pa_{--} S_{+++}(x)\,,\\
 \pa_{++} \bar{S}_{+--}(x) &=&\, - \pa_{--} \bar{S}_{+++}(x)\,,\\
 \pa_{++} T_{----}(x) &=&\, -\pa_{--} \Theta(x) \,,\\
 \pa_{++} \Theta(x) &=&\, - \pa_{--} T_{++++}(x) \,,
\eea
\esubeq
where we have defined as usual 
\bea
\Theta(x):=T_{++--}(x)=T_{--++}(x)
~.
\eea

To conclude this subsection, let us describe the ambiguity in the form of the $\cS$-multiplet which is parametrized by a choice of 
improvement term. 
If $\cU$ is a real superfield, we are free to modify the $\cS$-multiplet superfields as follows
\bsubeq    \label{S_mult_improvement}
\bea
    \cS_{\pm \pm} &\to \cS_{\pm \pm} + [ D_{\pm} , \Db_{\pm} ] \cU~ , \\
    \chi_{\pm} &\to \chi_{\pm} - \Db_+ \Db_- D_{\pm} \cU ~, \\
    \mathcal{Y}_{\pm} &\to \mathcal{Y}_{\pm} - D_{\pm} \Db_+ \Db_- \cU ~,
\eea
\esubeq
which keeps invariant the conservation equations \eqref{conservation-S}.
In general the $\cS$-multiplet is a reducible representation of supersymmetry and  some of its component can consistently be set 
to zero by a choice of improvement.
The reduced Ferrara-Zumino supercurrent multiplet, which plays a central role in our paper, is described next.

\subsection{The Ferrara-Zumino (FZ) multiplet and old-minimal supergravity}

If there exists a well-defined superfield $\cU$ such that $\chi_{\pm} = \Db_+ \Db_- D_{\pm} \cU$,  then we may 
use the transformation (\ref{S_mult_improvement}) to set $\chi_{\pm} = 0$ in the $\cS$-multiplet.  If in addition $k =  C^{(\pm)}=0$, then the fields $\mathcal{S}_{\pm \pm}$ and $\mathcal{Y}_{\pm}$ satisfy the defining equations for the Ferrara-Zumino (FZ) multiplet. In this case, it is conventional to rename the field $\mathcal{S}_{\pm \pm}$ to $\mathcal{J}_{\pm \pm}$ and write these defining equations as
\bsubeq    \label{FZ_constraint}
\bea
    \Db_{\pm} \mathcal{J}_{\mp \mp} &=& \pm \mathcal{Y}_{\mp}~ , \\
    D_{\pm} \mathcal{Y}_{\pm} &=& 0 ~, \\
    \Db_{\pm} \mathcal{Y}_{\mp} &=& 0~ , \\
    D_+ \mathcal{Y}_- + D_- \mathcal{Y}_+ &= &k' ~. 
\eea
\esubeq
The superfield $\mathcal{J}_{\pm \pm}$ in the FZ multiplet turns out to be associated to the axial $U(1)_A$ $R$-symmetry current, and satisfies the conservation equation
\be
\pa_{--}{\cal J}_{++}
-\pa_{++}{\cal J}_{--}
=0
~.
\ee
This multiplet, which has $4+4$ real components,
 is the dimensionally-reduced version of the $D=4$ $\cN=1$ FZ-multiplet \cite{Ferrara:1974pz}; 
 see Appendix \ref{components-S}\ for more details. All of the models we consider in section \ref{sec:models} have the property that 
 $\chi_{\pm}$ can be improved to zero;  that is, they all have a well-defined FZ-multiplet.

Just as the bosonic Hilbert stress tensor $T_{\mu \nu}$ represents the response function of the Lagrangian to a linearized perturbation
  $h_{\mu \nu}$ of the metric, the supercurrent multiplets correspond to linearized couplings to supergravity.\footnote{Rather than coupling to supergravity, one could define the supercurrent multiplets using a superspace Noether procedure, as is done for $D=4$ theories with $\, \mathcal{N} = 1$ supersymmetry in \cite{Magro:2001aj}. This was the approach followed for $(1,1)$ supersymmetry in \cite{Chang:2018dge}.}

  Different formulations of 
  off-shell supergravity couple to different supercurrent multiplets. If a theory has a well-defined FZ-multiplet, as is the case for all the 
  examples found in section \ref{sec:models}, then the theory can be consistently coupled to the old-minimal supergravity prepotentials 
  $H^{\pm \pm}$ and $\sigma$. The nomenclature ``old-minimal''  is again inherited from $D=4$ $\cN=1$ supergravity; see 
  \cite{Gates:1983nr,Buchbinder:1998qv} for pedagogical reviews and references. Here $H^{\pm \pm}$ is the conformal supergravity 
  prepotential---the analogue of the traceless part of the metric---and $\sigma$ is a chiral conformal compensator. 

We refer the reader to \cite{Grisaru:1994dm, Grisaru:1995dr, Grisaru:1995kn, Grisaru:1995py, Gates:1995du} 
and references therein for an exhaustive description of  $D=2$ $\,\cN=(2,2)$ off-shell supergravity in superspace, which we will use in 
our analysis; see also Appendix \ref{appendix:supercurrent_calculation}. For the scope of this work, it will be enough to know the 
structure of linearized old-minimal supergravity. For instance, at the linearized level the gauge symmetry of the supergravity 
prepotentials $H^{\pm\pm},\,\s$ and $\bar{\s}$, can be parameterized as follows
\bsubeq\label{old_minimal_gauge}
\bea
    \delta H^{++} &=& \frac{i}{2} \left( \Db_- L^+ - D_- \Lb^+ \right) ~,\\
    \delta H^{--} &= &\frac{i}{2} \left( \Db_+ L^- - D_+ \Lb^- \right) ~,\\
    \delta \sigma &=& - \frac{i}{2} \Db_+\Db_- \left( D_+ L^+ - D_- L^- \right) ~,\\
    \delta \bar{\sigma} &=& - \frac{i}{2} D_-D_+ \left( \Db_+ \Lb^+ - \Db_- \Lb^- \right) 
    ~,
\eea
\esubeq
in terms of  unconstrained spinor superfields $L^{\pm}$ and their complex conjugates. 

The conservation law (\ref{FZ_constraint}) for the FZ-multiplet can be derived by using the previous gauge transformations.
 The linearized supergravity couplings for a given model are written as%
 \footnote{We use the notation  $d^2\q:=d\q^- d\q^+$, $d^2\bar\q:=d\bar\q^+d\bar\q^-$ and  $d^4\q:=d^2\q d^2\qb$.}
\begin{align}
    \mathcal{L}_{\text{linear}} &= \int d^4 \theta \, \left( H^{++} \mathcal{J}_{++} + H^{--} \mathcal{J}_{--} \right)
    - \int d^2 \theta \, \sigma \, \mathcal{V}
        - \int d^2 \bar{\theta} \, \bar{\sigma} \, \overline{\mathcal{V}}
~,
\label{linearized_couplings_D_term}
\end{align}
with $\cV$ a chiral superfield and $\cVb$ its complex conjugate. Assuming the matter superfields satisfy their equations of motion, 
the change in the Lagrangian \eqref{linearized_couplings_D_term}  under the gauge transformation (\ref{old_minimal_gauge}) is
\begin{align}
    \delta \mathcal{L}_{\text{linear}} &= \int d^4 \theta \, \left( \delta H^{++} \mathcal{J}_{++} 
    + \delta H^{--} \mathcal{J}_{--} 
\right)
   - \int d^2 \theta \, \d\sigma \, \mathcal{V}
        - \int d^2 \bar{\theta} \, \d\bar{\sigma} \, \overline{\mathcal{V}}
     \nonumber \\
    &= \frac{i}{2} \int d^4 \theta \, \Big\{ \left( \Db_- L^+ - D_- \Lb^+ \right) \mathcal{J}_{++} 
    + \left( \Db_+ L^- - D_+ \Lb^- \right) \mathcal{J}_{--} \nonumber \\
    &\qquad~~~~~~~~~~~~
     -  \left( D_+ L^+ - D_- L^- \right) \mathcal{V} 
    -\left( \Db_+ \Lb^+ - \Db_- \Lb^- \right) \overline{\mathcal{V}} \Big\} \nonumber \\
    &= \frac{i}{2} \int d^4 \theta \, \Big\{ L^+ \left( \Db_- \mathcal{J}_{++} +  D_+ \mathcal{V}  \right) 
    + L^- \left( \Db_+ \mathcal{J}_{--} -  D_- \mathcal{V}  \right) + \text{c.c.} \Big\} ~, 
\end{align}
where we have integrated by parts. Demanding that the variation vanishes for any gauge parameter $L^{\pm}$ gives
\be
    \Db_{-} \mathcal{J}_{++} + D_+ \mathcal{V} = 0 ~,~~~~~~
    \Db_+ \mathcal{J}_{--} - D_- \mathcal{V}  = 0 
    ~.
    \label{FZ-2}
\ee
This matches the constraints (\ref{FZ_constraint}) for the FZ-multiplet if we identify
\begin{align}
    \mathcal{Y}_{\pm} = D_{\pm} \mathcal{V} 
         ~,
\end{align}
and set $k'= 0$.

As we will soon see, studying $T \Tb$ deformations requires consideration of a 
composite operator constructed out of the square of the supercurrent multiplet. 
Hence to solve the $T \Tb$ flow equations we need to be able to calculate the supercurrent multiplet explicitly. 
The coupling to supergravity provides a straightforward prescription for computing the FZ-multiplet for matter models 
that can be coupled to old-minimal supergravity.%
\footnote{Though we will not need it in our paper, it is worth mentioning that the non-minimal supergravity results of 
\cite{Grisaru:1994dm,Grisaru:1995dr,Grisaru:1995kn,Grisaru:1995py,Gates:1995du} allow the computation of the supercurrent 
multiplet for more general classes of models.} 
In particular, for a given $\cN=(2,2)$ matter theory we will:
\begin{enumerate}

    \item Begin with an undeformed superspace Lagrangian  $\mathcal{L}$ in flat $\cN=(2,2)$ superspace.
    
    \item Minimally couple $\mathcal{L}$ to the supergravity superfield prepotentials $H^{\pm\pm}$, $\sigma$ and  $\bar\sigma$.
    
    \item Extract the superfields $\mathcal{J}^{\pm\pm}$, $\mathcal{V}$ and  $\overline{\mathcal{V}}$
     which couple linearly to $H^{\pm\pm}$,
    $\sigma$ and  $\bar\sigma$, respectively, in the D- 
    and F-terms of (\ref{linearized_couplings_D_term}).
    
\end{enumerate}
Thanks to the analysis given above, the superfields $\mathcal{J}^{\pm\pm}$, $\mathcal{V}$ and  $\overline{\mathcal{V}}$ 
will automatically satisfy the FZ-multiplet constraints \eqref{FZ-2}.
A detailed description of the computation of the FZ-multiplet for the models relevant for our paper is given in Appendix 
\ref{appendix:supercurrent_calculation}.


\section{The \texorpdfstring{$T \Tb$}{T Tbar} Operator and \texorpdfstring{$\cN = (2,2)$}{N=(2,2)} Supersymmetry}
 \label{secTTbar}

After having reviewed in the previous section the structure of the  $\cS$-multiplet, we are 
ready to describe $\cN=(2,2)$ $T \Tb$ deformations.

\subsection{The \texorpdfstring{$\calT \calTb$}{super T Tbar} operator}

Given a $D=2$ $\cN=(2,2)$ supersymmetric theory with an $\cS$-multiplet, 
we define the supercurrent-squared deformation of this theory, denoted $\calT \calTb$ in analogy with $T \Tb$, by the flow equation
\begin{align}
    \partial_\lambda \mathcal{L} &= - \frac{1}{8} \calT \calTb 
    ~,
\end{align}
where $\calT \calTb$ is constructed from current bilinears with
\begin{align}
     \calT \calTb &\equiv 
     -\int d^4 \theta \, \mathcal{S}_{++} \mathcal{S}_{--}
    - \left( \int d \theta^-  d \theta^+ \, \chi_+ \chi_- + \int d \thetab^-  d \theta^+ \, \calYb_+ \calY_- + \text{c.c.} \right) ~,
    \label{TTbar-S}
\end{align}
and where the factor of $\frac{1}{8}$ is chosen for later convenience.
This deformation generalizes the results we recently obtained for $D=2$ theories possessing $\cN=(0,1)$, $\cN=(1,1)$ and 
$\cN=(0,2)$ supersymmetry \cite{Chang:2018dge, Baggio:2018rpv, Jiang:2019hux} to theories with  $\cN=(2,2)$ supersymmetry.

Let us recall the form of the $T \Tb$ composite operator \cite{Zamolodchikov:2004ce}, which we denote
\bea
T \Tb(x)=T_{++++}(x)\,T_{----}(x)-\big[\Theta(x)\big]^2
~.
\label{component-TTbar}
\eea
An important property of the $\cN=(0,1)$, $\cN=(1,1)$ and $\cN=(0,2)$ cases is that the $T \Tb$ operator turns out to be
 the bottom component of a long supersymmetric multiplet. This is true up to both total vector derivatives ($\pa_{++}$ and $\pa_{--}$), 
 and terms that vanish upon using the supercurrent conservation equations (Ward identities). For this reason, in the supersymmetric 
 cases studied previously, the original $T \Tb$ deformation of \cite{Zamolodchikov:2004ce} is manifestly supersymmetric and 
 equivalent to the deformations constructed in terms of the full superspace integrals of primary supercurrent-squared composite 
 operators \cite{Chang:2018dge, Baggio:2018rpv, Jiang:2019hux}. 

Remarkably, despite the much more involved structure of the $(2,2)$ $\cS$-multiplet compared to theories with fewer 
supersymmetries, it is possible to prove that the following relation holds:
\begin{align}
\calT\calTb(x)
=
T \Tb(x)
+{\rm EOM's}
+\pa_{++}(\cdots)
+\pa_{--}(\cdots)
~.
\label{calTTb=TTb}
\end{align}
In~\eqref{calTTb=TTb}, we  use ${\rm EOM's}$ to denote terms that are identically zero when \eqref{conservation-S} are used.
Showing~\eqref{calTTb=TTb} requires using  \eqref{comp-S}--\eqref{comp-Y}, along with several cancellations, integration by parts 
and the use of the $(2,2)$ $\cS$-multiplet conservation equations
\eqref{conservation-S}. 

In fact, the specific combination of current superfields given in \eqref{TTbar-S} was chosen precisely for \eqref{calTTb=TTb} to hold. 
The combination \eqref{calTTb=TTb} is also singled out by being invariant under the improvement transformation 
\eqref{S_mult_improvement}.
The important implication of  \eqref{calTTb=TTb} is that the $T \Tb$ deformation for an $\cN=(2,2)$ supersymmetric quantum field 
theory is manifestly supersymmetric and equivalent to the $\calT \calTb$ deformation of eq.~\eqref{TTbar-S}. 

Note that in the $\cN=(2,2)$ case the deformation we have introduced in \eqref{TTbar-S} is conceptually different from the cases 
with less supersymmetry. Specifically, the deformation is not given by the descendant of a \emph{single} composite superfield. 
On the other hand, suppose the $\cS$-multiplet is such that $C^{(\pm)}=k=k'=0$ and it is possible to improve the superfields  
$\chi_\pm$ and $\cY_{\pm}$ to a case where
\bsubeq\label{simplified-S-multiplet}
\bea
&
\cY_\pm=D_\pm\cV
~,~~~
\overline{\cY}_\pm=\Db_\pm\overline{\cV}
~,
\\
&
\chi_+=i\Db_+\overline{\cal B}
~,~~~
\chi_-=i \Db_-{\cal B}
~,~~~
\bar{\chi}_+= - i D_+{\cal B}
~,~~~
\bar{\chi}_-= - i D_-\overline{\cal B}
~,
\eea
\esubeq
with $\cV$ chiral and ${\cal B}$ twisted-chiral:
\bsubeq
\bea
&\Db_\pm \cV=0~,~~~
D_\pm \overline{\cal V}=0
~,
\\
&\Db_+ {\cal B}=D_-{\cal B}=0~,~~~
D_+ \overline{\cal B}=\Db_-\overline{\cal B}=0
~.
\eea
\esubeq
In this case \eqref{TTbar-S} simplifies to
\bea
\calT\calTb
&=&  - \int d^4 \theta \, \mathcal{S}_{++} \mathcal{S}_{--} 
+ \left( \int d \theta^-  d \theta^+ \, \Db_+\overline{\cal B}\Db_-\overline{\cal B}
- \int d \thetab^- \, d \theta^+ \, \Db_+ \overline{\mathcal{V}} D_- \mathcal{V} + \text{c.c.} \right) 
\non \\
        &=& -\int d^4 \theta \, \left( 
        \mathcal{S}_{++} \mathcal{S}_{--} 
        - 2 \mathcal{B} \overline{\mathcal{B}}
        - 2 \mathcal{V} \overline{\mathcal{V}} \right) 
        ~,
        \label{simplifiedTTbar}
\eea
and we see that, up to EOM's,  $T \Tb(x)$ is the bottom component of a long supersymmetric multiplet. 
In this situation, once we define the composite superfield 
\bea
\cO(\z):=
- \mathcal{S}_{++}(\z) \mathcal{S}_{--}(\z)
+ 2 \mathcal{B}(\z) \overline{\mathcal{B}}(\z)
+ 2 \mathcal{V}(\z) \overline{\mathcal{V}}(\z)
~,
\label{primary-TTbar}
\eea
eq.~\eqref{calTTb=TTb} turns into the equivalent result%
\footnote{In the subsequent discussion by $\q=0$ we will always mean $\q^\pm=\qb^{\pm}=0$.}
\begin{align}
 \int d^4 \theta\,\cO (\z)
 =D_-D_+\Db_+\Db_-\cO(\z)|_{\q=0}
=
T\Tb(x)
+{\rm EOM's}
+\pa_{++}(\cdots)
+\pa_{--}(\cdots)
~,
\label{calTTb=TTb-2}
\end{align}
stating that the D-term of the operator $\cO(\z)$ is equivalent to the standard $T\Tb(x)$ operator. 

For a matter theory that can be coupled to old-minimal supergravity, leading to the FZ-multiplet described by \eqref{FZ-2},
 the operator $\cO(\z)$ further simplifies thanks to the fact that the twisted-(anti-)chiral operators ${\cal B}$ and $\overline{{\cal B}}$ 
 disappear. For these cases, the $T \Tb$ flow turns into the following equation
\be
        \partial_{\lambda} \mathcal{L} 
  = \frac{1}{8} \int d^4 \theta \, \left( \mathcal{J}_{++} \mathcal{J}_{--} - 2 \mathcal{V} \overline{\mathcal{V}} \right) 
  ~.
  \label{FZTTbar}
  \ee
This will be our starting point in analyzing  $\cN=(2,2)$ deformed models in section \ref{sec:models}.


\subsection{Point-splitting and well-definedness}
\label{point-plitting-section}

The $T \Tb (x)$ operator \eqref{component-TTbar} is quite magical because it is a well-defined irrelevant composite local operator, 
free of short distance divergences \cite{Zamolodchikov:2004ce}. In fact,  this property generalizes to the larger class of operators 
\begin{align}
[A_{s}(x)\, A'_{s'}(x)-B_{s+2}(x)\, B_{s'-2}(x)]
\end{align} 
where $(A_s,B_{s+2})$ and $(A'_{s'},B'_{s'-2})$ are two pairs of conserved currents with spins $s$ and $s'$. The operator 
$T \Tb (x)$ is a particular example with $s=s'=0$. As  proven in \cite{Smirnov:2016lqw}, these composite operators of 
``Smirnov-Zamolodchikov''-type have a well-defined point splitting which is free of short-distance divergences. In the case of 
$\cN = (0,1)$ and $\cN = (1, 1)$ supersymmetric $T\Tb$ deformations, the entire supermultiplet whose bottom component is 
$T \Tb (x)$ is comprised of well-defined  Smirnov-Zamolodchikov-type operators \cite{Baggio:2018rpv,Chang:2018dge}. In the 
$\cN=(0,2)$ case, the primary%
\footnote{We denote as primary operator the top component of a supersymmetric multiplet even when the theory is not 
superconformal.} 
operator whose bottom component is $T \Tb(x)$ is not of Smirnov-Zamolodchikov-type. Nevertheless, also in this case it was recently 
shown that, thanks to supersymmetry, the whole multiplet is  well-defined \cite{Jiang:2019hux}.

In the $\cN=(2,2)$ case it is clear that the situation is more complicated than any of the cases mentioned above. 
First, in the general situation, according to \eqref{TTbar-S}, the $\calT\calTb$ deformation is a linear combination of a D-term together 
with chiral and twisted-chiral F-terms contributions. Though the F-terms might be protected by standard perturbative 
non-renormalization theorems (see, for example, \cite{Gates:1983nr,Buchbinder:1998qv} for the $D=4$ $\cN=1$ case which 
dimensionally reduces to  $D=2$ $\cN=(2,2)$), the D-term associated to the $\cS_{++}\cS_{--}$ operator has no clear reason to be 
protected in general from short-distance  divergences in point-splitting regularization, and hence has no obvious reason to be 
well-defined. This indicates that there might be a clash between supersymmetry and a point-splitting procedure in the general setting. 

We will not attempt to analyze this issue in full generality in the current paper; instead our aim is to describe a subclass of models 
for which the $\calT \calTb$ deformation turns out to be well-defined. A natural restriction to impose is that the $\cS$-multiplet is 
constrained by \eqref{simplified-S-multiplet} and the $\calT\calTb$ deformation is therefore described by the D-term 
\eqref{simplifiedTTbar}. By trivially extending the arguments used in \cite{Jiang:2019hux} for the $\cN=(0,2)$ case, it is not difficult to 
show that these restrictions are sufficient to imply that the multiplet described by the $\cN=(2,2)$ primary operator $\cO(\z)$ of 
\eqref{primary-TTbar} is indeed well-defined despite not being of Smirnov-Zamolodchikov-type. As in the $\cN=(0,2)$,
unbroken $\cN=(2,2)$ supersymmetry turns out to be the reason for this to happen.

Let us quickly explain how this works for the FZ-multiplet and the deformation \eqref{FZTTbar}, which are the main players in our 
paper. Note, however, that the same argument extends to more general cases where both chiral and twisted-chiral 
current superfields, $\chi_{\pm}$ and $\cY_{\pm}$,
satisfying \eqref{simplified-S-multiplet}
are turned on. 
We also refer to  \cite{Jiang:2019hux} for details that we will skip in the following discussion, which are trivial extensions from the 
$(0,2)$ to the $(2,2)$ case.

A first indication of the well-definedness of the multiplet associated to $\cO(\z)$ comes by looking  at the vacuum expectation 
value of its lowest component. Define the primary composite operator
\bea
O(x):=-j_{--}(x)j_{++}(x)+2v(x)\bar{v}(x)
=  \cO(\z)|_{\q=0}
~,
\label{comp-primary}
\eea
and its point-split version
\bea
O(x,x'):=
-j_{--}(x)j_{++}(x')
+v(x)\bar{v}(x')
+\bar{v}(x)v(x')
~,
\label{comp-primary-ps}
\eea
where
\bea
j_{\pm\pm}(x):={\cal J}_{\pm\pm}(\z)|_{\q=0}
~,~~~
v(x):=\cV(\z)|_{\q=0}
~,~~~
\bar{v}(x):=\overline{\cV}(\z)|_{\q=0}
~.
\label{FZ-3}
\eea
Note that equation \eqref{FZ-2} implies the following relation among the component operators
\bea
\big[\overline{Q}_\pm,j_{\mp\mp}(x)\big]
=\pm\big[Q_\mp,v(x)\big]
~,~~~
\big[Q_\pm,j_{\mp\mp}(x)\big]
=\pm\big[\overline{Q}_\mp,\bar{v}(x)\big]
~,
\label{conservation-components}
\eea
with $Q_{\pm}$ and $\overline{Q}_\pm$ denoting the $\cN=(2,2)$ supercharges.%
\footnote{Given an operator $F(x)$ defined as the $\theta=0$ component of the superfield $\cF(\zeta)$, $F(x):=\cF(\z)|_{\theta=0}$, 
then its supersymmetry transformations are such that $\big[Q_\pm,F(x)\big\}=\cQ_\pm\cF(\z)|_{\theta=0}
=D_\pm\cF(\z)|_{\theta=0}$
and
$\big[\overline{Q}_\pm,F(x)\big\}
=\overline{\cQ}_\pm\cF(\z)|_{\theta=0}
=\overline{D}_\pm\cF(\z)|_{\theta=0}$.
}
By then using $\pa_{\pm\pm}=i\{Q_\pm,\overline{Q}_\pm\}$, $\{Q_+,Q_-\}=\{\overline{Q}_+,\overline{Q}_-\}=0$, 
$[\overline{Q}_\pm,v(x)]=[Q_\pm,\bar{v}(x)]=0$, 
super-Jacobi identities,
together with the conservation equations \eqref{conservation-components}, 
and the assumption that the vacuum is invariant under  supersymmetry, it is straightforward to show that vacuum expectation value
 of $O(x,x')$ satisfies
\bea
\pa_{++}\left\langle j_{--}(x)j_{++}(x')\right\rangle 
&=&
i\left\langle
\left[\{Q_+,\overline{Q}_+\},j_{--}(x)\right]
j_{++}(x')\right\rangle 
\non\\
&=&
i\left\langle
\left\{Q_+,[Q_-,v(x)]\right\}
j_{++}(x')
+\left\{\overline{Q}_+,[\overline{Q}_-,\bar{v}(x)]\right\}
j_{++}(x')
\right\rangle 
\non\\
&=&
-i\left\langle
[Q_+,v(x)]
[Q_-,j_{++}(x')]
+[\overline{Q}_+,\bar{v}(x)]
[\overline{Q}_-,j_{++}(x')]
\right\rangle 
\non\\
&=&
i\left\langle
[Q_+,v(x)]
[\overline{Q}_+,\bar{v}(x')]
+[\overline{Q}_+,\bar{v}(x)]
[Q_+,v(x')]
\right\rangle 
\non\\
&=&
\left\langle
\left[i\{\overline{Q}_+,Q_+\},v(x)\right\}
\bar{v}(x')
+\left[i\{Q_+,\overline{Q}_+\},v(x)\right\}
\bar{v}(x')
\right\rangle 
\non\\
&=&
\pa_{++}\left\langle
v(x)\bar{v}(x')+v(x)\bar{v}(x')
\right\rangle 
~,
\eea
and, after performing a similar calculation for
$\left\langle \pa_{--}j_{--}(x)j_{++}(x')\right\rangle
=-\left\langle j_{--}(x)\pa'_{--}j_{++}(x')\right\rangle $,
it is clear that the relation
\be
\pa_{\pm\pm}\left\langle O(x,x')\right\rangle 
=0
\ee
holds. 
Therefore, $\left\langle O(x,x')\right\rangle $ is independent of the positions and free of short distance divergences. 
It is worth noting that similarly to the argument showing that the two point function of two chiral or twisted-chiral operators is 
independent of the positions $x$ and $x'$, the previous analysis for $\left\langle O(x,x')\right\rangle$ necessarily relies on unbroken 
$\cN=(2,2)$ supersymmetry.

The argument given above can be generalized to a statement about operators in  superspace  in complete analogy to the 
$\cN=(0,2)$ case of \cite{Jiang:2019hux}. Let us investigate the short distance singularities in the bosonic coordinates by defining  a 
point-split version of the  $\cN=(2,2)$ primary $\calT\calTb$ operator,
 \bea
 \cO(x,x',\q)
 :=
 - {\cal J}_{--}(x,\q)\,{\cal J}_{++}(x',\q)   
 +\cV(x,\q)\, \overline{\cV}(x',\q)
 +\overline{\cV}(x,\q)\, \cV(x',\q)
 ~.
 \label{pointsplit-primary}
 \eea
We want to show that the preceding bilocal superfield is free of short distance divergences in the limit $x\to x'$. 
A straightforward calculation shows that
\bea\hspace{-12pt}
\pa_{++}\cO(x,x',\q)
&=&
- \Big\{
 i  D_+\cV(\z)\left[D'_-\cJ_{++}(\z')+\Db'_+\overline{\cV}(\z')\right]
+ i   \Db_+\overline{\cV}(\z)\left[\Db'_-\cJ_{++}(\z')+D'_+\cV(\z')\right]
\non\\
&&
+ i  (\cQ_++ \cQ'_+)
\left[(\Db_+\overline{\cV}(\z))\cV(\z')\right]
+ i  (\overline{\cQ}_++\overline{\cQ}'_+)
\left[(D_+\cV(\z))\overline{\cV}(\z')\right]
\non\\
&&
+ i  (\cQ_-+\cQ'_-)
\left[(D_+\cV(\z))\cJ_{++}(\z')\right]
+ i   (\overline{\cQ}_-+\overline{\cQ}'_-)
\left[(\Db_+\overline{\cV}(\z))\cJ_{++}(\z')\right]
\non\\
&&
+(\pa_{++}+\pa'_{++})
\left[\qb^+(\Db_+\overline{\cV}(\z))\cV(\z')
+\qb^-(D_+\cV(\z))\cJ_{++}(\z')\right]
\non\\
&&
-(\pa_{++}+\pa'_{++})
\left[\q^+ (D_+\cV(\z))\overline{\cV}(\z')
+\q^-(\Db_+\overline{\cV}(\z))\cJ_{++}(\z')\right]
\Big\}\Big|_{\q=\q'}
~.
\label{coinciding-trick}
\eea
Note that the first line in the preceding expression 
is zero because of the FZ conservation equations \eqref{FZ-2}, which hold up to contact terms in correlation functions. 
The other lines are either total vector derivatives or 
supersymmetry transformations of bilocal operators. A 
similar equation holds for $\pa_{--}\cO(x,x',\q)$ showing 
that the operator $\cO(x,x',\q)$ satisfies
\bea
\pa_{\pm\pm}\cO(x,x',\q)
&=&
0
\,+\,\textrm{EOM's}
\,+\,[P,\cdots]
\,+\,[ Q,\cdots]
~,
\label{paO}
\eea
where $[P,\cdots]$ and $[ Q,\cdots]$ schematically indicate a translation and supersymmetry transformation of some bilocal superfield 
operator.%
\footnote{See Appendix A of \cite{Jiang:2019hux} for the relation between the operators $(\cQ_\pm+ \cQ'_\pm)$, 
$(\overline{\cQ}_\pm+ \overline{\cQ}'_\pm)$ and the generators of supersymmetry transformations on bilocal superfields such as 
$\cO(x,x',\q)$. The extension of that analysis from $\cN=(0,2)$ to $\cN=(2,2)$ is straightforward.} 
To conclude, by employing an OPE argument completely analogous to the one originally given by Zamolodchikov in 
\cite{Zamolodchikov:2004ce} and extended to the $\cN=(0,2)$ supersymmetric case in \cite{Jiang:2019hux}, one can show that 
eq.~\eqref{paO} implies
\be
\cO(x,x',\q)
=
\cO(\z)
\,+\,
{\rm derivative~terms}
~.
\ee 
Here ``derivative terms'' indicate superspace covariant derivatives 
$D_A=(\pa_{\pm\pm},D_\pm,\Db_\pm)$ 
acting on local superfield operators while $\cO(\z)$ arises from the regular, non-derivative part of the OPE of $\cO(x,x',\q)$. 
As a result the integrated operator 
\bea
S_{\cO} 
= 
\int d^2x\,d^4\q\,
\lim_{\varepsilon\to0}\cO(x, x+\varepsilon, \theta) 
=\int d^2x\,d^4\q:\cO(x, x, \theta):
~,
\eea
which can be considered as a definition of the integrated
$\calT\calTb(x)$ operator,%
\footnote{Note that, consistently, one can show that
\be
\{Q_+,[\overline{Q}_+,\{Q_-,[\overline{Q}_-,O(x,x')]\}]\}
=T_{----}(x)T_{++++}(x')
-\Theta(x)\Theta(x')
+\textrm{EOM's}
+[P,\cdots]
\ee
implying
that the descendant of the point-split
primary operator $O(x)$  is equivalent, up to Ward identities and total vector derivatives ($\pa_{\pm\pm}$), to
the point-split version of the descendant 
$T\Tb(x)$ operator.}
is free of short distance divergences and well-defined in complete analogy to the non-supersymmetric case
 \cite{Zamolodchikov:2004ce} and the $\cN=(0,1)$, $\cN=(1,1)$, and $\cN=(0,2)$  cases 
 \cite{Baggio:2018rpv, Chang:2018dge, Jiang:2019hux}.


\section{Deformed \texorpdfstring{$(2,2)$}{(2,2)} Models } \label{sec:models}

In this section, we will apply our supercurrent-squared deformation (\ref{FZTTbar}) to a few examples of
 $\mathcal{N} = (2,2)$ supersymmetric theories for a chiral multiplet $\Phi$. The superfield $\Phi$ can be written in components as
\bea
    \Phi &=& \phi  
    + \theta^+ \psi_+ 
    + \theta^- \psi_-
    + \theta^+ \theta^- F
    - i \theta^+ \thetab^+ \partial_{++} \phi 
    - i \theta^- \thetab^- \partial_{--} \phi 
     \non\\
    &&
    - i \theta^+ \theta^- \thetab^- \partial_{--} \psi_+ 
     - i \theta^- \theta^+ \thetab^+ \partial_{++} \psi_- 
    - \theta^+ \theta^- \thetab^- \thetab^+ \partial_{++} \partial_{--} \phi
    ~,
\eea
where $\phi$ is a complex scalar field, $\psi_{\pm}$ are Dirac fermions, and $F$ is a complex auxiliary field. 
The multiplet $\Phi$ satisfies the chirality constraint $\Db_{\pm} \Phi = 0$.

We denote the physical Lagrangian by $\mathcal{L}$ and the superspace D-term Lagrangian by $\mathcal{A}$, so that
\begin{align}
    S = \int d^2 x \, \mathcal{L} = \int d^2 x \, d^4 \theta \, \mathcal{A} ~.
\end{align}
A broad class of 
two-derivative theories for a chiral superfield can be described by superspace Lagrangians of the form
\begin{align}
    \mathcal{L} = \int d^4 \theta \, K ( \Phi , \Phib) + \int d^2 \theta \, W ( \Phi ) + \int d^2 \thetab \Wb ( \Phib ) 
    ~,
\end{align}
where $K ( \Phi , \Phib )$ is a real function called the K\"ahler potential and $W ( \Phi )$ is a holomorphic function called the 
superpotential. These are $\cN=(2,2)$ Landau-Ginzburg models. In order for the kinetic terms of the component fields of $\Phi$ to 
have the correct sign, we will assume that $K_{\Phi \Phib} = \frac{\partial^2 K}{\partial \Phi \partial \Phib}$ is positive.  

Although we will not expand on this point in detail, all the results found in this section can be derived almost identically for the case of 
a generic model of a single scalar twisted-chiral superfield $\calY$, 
$\Db_+\calY=D_-\calY=0$,
and its conjugate. This is not surprising since theories containing only chiral superfields are physically equivalent to theories 
formulated in terms of twisted-chiral superfields; see,
 for example,~\cite{Grisaru:1994dm,Grisaru:1995dr,Grisaru:1995kn,Grisaru:1995py,Gates:1995du} 
 for a discussion of this equivalence in models with global and local supersymmetry. There are also many more involved $(2,2)$ 
 theories that one might also want to study involving chiral, twisted-chiral and semi-chiral superfields; see,
  for example,~\cite{Caldeira:2018ynv} for a recent discussion and references. 
  For this analysis,  we have chosen to consider only models based on a single 
chiral multiplet.

\subsection{K\"ahler potential} \label{subsec:kahler}

First we will set the superpotential $W$ to zero and begin with an undeformed superspace Lagrangian of the form
\begin{align}
    \mathcal{L} = \int d^4 \theta \, K ( \Phi , \Phib )
\end{align}
for some K\"ahler potential $K$. To leading order around this undeformed theory, the FZ supercurrents are
\bsubeq
\bea
    \mathcal{J}_{\pm \pm} &=& 2 K_{\Phi \Phib} D_{\pm} \Phi \Db_{\pm} \Phib~ , \\
    \mathcal{V} &=& 0~ ,
\eea
\esubeq
where $K_{\Phi} = \frac{\partial K}{\partial \Phi}, K_{\Phi \Phib} = \frac{\partial^2 K}{\partial \Phi \partial \Phib}$, etc. 
Therefore, at first order the supercurrent-squared deformation driven by 
$\cO=
\left(- \mathcal{S}_{++} \mathcal{S}_{--}
+ 2 \mathcal{V} \overline{\mathcal{V}}\right)$
will source a four-fermion contribution in the D-term, giving
\begin{align}
    \mathcal{L}^{(1)} = \mathcal{L}^{(0)} + \frac{1}{2} \lambda K_{\Phi \Phib}^2 D_+ \Phi \Db_+ \Phib D_- \Phi \Db_- \Phib ~.
\end{align}
Next, we would like to find the all-orders solution for the deformed theory. We make the ansatz that, at finite deformation parameter 
$\lambda$, the Lagrangian takes the form
\begin{align}
    \mathcal{L}_{\lambda} = \int d^4 \theta  \Big\{ K ( \Phi, \Phib ) 
    + f ( \lambda, x, \xb, y ) K_{\Phi \Phib}^2 D_+ \Phi \Db_+ \Phib D_- \Phi \Db_- \Phib \Big\}~ , 
    \label{kahler_ansatz}
\end{align}
where we define the combinations
\begin{align}
    x = K_{\Phi \Phib} \partial_{++} \Phi \partial_{--} \Phib
    ~, \qquad
    y = K_{\Phi \Phib} \left( D_+ D_- \Phi \right) \left( \Db_+ \Db_- \Phib \right)
    ~.
\end{align}
Using the results in Appendix 
\ref{appendix:supercurrent_calculation}, one finds that the 
superfields $\mathcal{J}_{\pm \pm}$ and $\mathcal{V}$ appearing in 
our supercurrent-squared deformation, computed for the Lagrangian
\eqref{kahler_ansatz},
are given by
\begin{align} \hspace*{-50pt}
    \mathcal{J}_{++} &= 2 K_{\Phi \Phib} D_+ \Phi \Db_+ \Phib 
    \left[ 1 + f ( x + \xb - 3 y ) 
    + x \frac{\partial f}{\partial x} ( \xb - y ) 
    + \xb \frac{\partial f}{\partial \xb} ( x - y ) 
    + y \frac{\partial f}{\partial y} ( x + \xb - 2 y ) \right]
     \nonumber \\
    &\quad 
    + 2 K_{\Phi \Phib}^2 D_- \Phi \Db_- \Phib \partial_{++} \Phi \partial_{++} \Phib
     \left[ - f 
     - x \frac{\partial f}{\partial x} 
     - \xb \frac{\partial f}{\partial \xb} 
     + y \left( \frac{\partial f}{\partial x} 
     + \frac{\partial f}{\partial \xb} \right) \right] 
      \nonumber \\
    &\quad -2 i K_{\Phi \Phib}^2 D_+ \Phi D_- \Phi \partial_{++} \Phib \Db_+ \Db_- \Phib
     \left[ - f  
     + ( x - \xb ) \frac{\partial f}{\partial \xb} 
     + (x-y) \frac{\partial f}{\partial y}  \right]  
     \nonumber \\
    &\quad 
    - 2 i K_{\Phi \Phib}^2 \Db_+ \Phib \Db_- \Phib \partial_{++} \Phi D_+ D_- \Phi 
    \left[ f 
    + ( x - \xb ) \frac{\partial f}{\partial x} 
    + ( y - \xb ) \frac{\partial f}{\partial y}  \right]   
    ~,
    \label{kahler_jpp}
\end{align}
and
\begin{align} \hspace*{-50pt}
    \mathcal{J}_{--} &= 2 K_{\Phi \Phib} D_- \Phi \Db_- \Phib 
    \left[ 1 
    + f ( x + \xb - 3 y ) 
    + x \frac{\partial f}{\partial x} ( \xb - y ) 
    + \xb \frac{\partial f}{\partial \xb} ( x - y ) 
    + y \frac{\partial f}{\partial y} ( x + \xb - 2 y ) \right] 
    \nonumber \\
    &\quad
    +2 K_{\Phi \Phib}^2 D_+ \Phi \Db_+ \Phib \partial_{--} \Phi \partial_{--} \Phib 
    \left[ - f
     - x \frac{\partial f}{\partial x}
      - \xb \frac{\partial f}{\partial \xb} 
      + y \left( \frac{\partial f}{\partial x} 
      + \frac{\partial f}{\partial \xb} \right) \right]
       \nonumber \\
    &\quad 
     - 2 i K_{\Phi \Phib}^2 D_+ \Phi D_- \Phi \partial_{--} \Phib  \Db_+ \Db_- \Phib 
     \left[ - f 
     + ( \xb - x ) \frac{\partial f}{\partial x} 
     + ( \xb - y ) \frac{\partial f}{\partial y} \right] 
      \nonumber \\
    &\quad 
    - 2 i K_{\Phi \Phib}^2 \Db_+ \Phib \Db_- \Phib \partial_{--} \Phi  D_+ D_- \Phi 
    \left[ f 
    + ( \xb - x ) \frac{\partial f}{\partial \xb} 
    + (y - x ) \frac{\partial f}{\partial y} \right]~  ,
    \label{kahler_jmm}
\end{align}
and
\begin{align}\hspace*{-25pt}
    \mathcal{V} &= 2 K_{\Phi \Phib}^2
     \left( f 
     + y \frac{\partial f}{\partial y} 
     + x \frac{\partial f}{\partial x} 
     + \xb \frac{\partial f}{\partial \xb} \right) 
     \Big[ - i \partial_{++} \Phib ( D_+ D_- \Phi ) D_- \Phi \Db_- \Phib 
     + \partial_{++} \Phib \partial_{--} \Phib D_- \Phi D_+ \Phi 
     \nonumber \\
    &\quad 
    - \Db_- \Phib \Db_+ \Phib \left( D_+ D_- \Phi \right)^2 
    - i \partial_{--} \Phib ( D_+ D_- \Phi ) D_+ \Phi \Db_+ \Phib 
    \Big] 
    ~.
    \label{kahler_v}
\end{align}
The supercurrent-squared flow then induces a differential equation for the superspace Lagrangian 
$\mathcal{A}_\lambda$ (where, again, $\mathcal{L}_\lambda = \int d^4 \theta \, \mathcal{A}_\lambda$) given by
\begin{align}
    \frac{d}{d \lambda} \mathcal{A}_\lambda &=- \frac{1}{8}\cO= 
    \frac{1}{8} \left( \mathcal{J}_{++} \mathcal{J}_{--} 
    - 2 \mathcal{V} \overline{\mathcal{V}} \right)~ .
    \label{sc2_kahler_flow}
\end{align}
Given our ansatz (\ref{kahler_ansatz}), we see that
\begin{align}
    \frac{d \mathcal{A}_\lambda}{d \lambda} = \frac{df}{d \lambda} K_{\Phi \Phib}^2 \, D_+ \Phi \Db_+ \Phib D_- \Phi \Db_- \Phib ~.
\end{align}
On the other hand, plugging in our expressions (\ref{kahler_jpp}), (\ref{kahler_jmm}), and (\ref{kahler_v}) 
for the supercurrents into the right hand side of (\ref{sc2_kahler_flow}) also gives a result proportional to 
$K_{\Phi \Phib}^2 D_+ \Phi \Db_+ \Phib D_- \Phi \Db_- \Phib$. Equating the coefficients, we find a differential equation for $f$:
\begin{align}\hspace{-15pt}
    \frac{d f}{d \lambda} 
    =&\, 
    \frac{1}{2} \Bigg\{ - \xb y \left[ f + ( \xb - x ) \frac{\partial f}{\partial \xb} 
    + (y - x) \frac{\partial f}{\partial y }  \right]^2 
    - x y \left[ f + ( x - \xb ) \frac{\partial f}{\partial x} 
    + ( y - \xb ) \frac{\partial f}{\partial y} \right]^2 
     \nonumber \\
    & 
    + 2 ( x - y ) ( y - \xb ) \left[ f + y \frac{\partial f}{\partial y} + \xb \frac{\partial f}{\partial \xb}
     + x \frac{\partial f}{\partial x} \right]^2 
     + x \xb \left[ f + ( \xb - y ) \frac{\partial f}{\partial \xb} 
     + ( x - y ) \frac{\partial f}{\partial x} \right]^2
      \nonumber \\
    &+ \left[ 1 + ( x + \xb - 3 y ) f + ( x + \xb - 2 y ) y \frac{\partial f}{\partial y} + \xb ( x - y ) \frac{\partial f}{\partial \xb} 
    + x ( \xb - y ) \frac{\partial f}{\partial x} \right]^2 \Bigg\} 
    ~.
    \label{bg_def_unsimp}
\end{align}
In particular, this shows that our ansatz (\ref{kahler_ansatz}) for the finite-$\lambda$ superspace action is consistent: 
the supercurrent-squared deformation closes on an action of this form. It could have been otherwise: the flow equation might have
 sourced additional terms proportional, say, to two-fermion combinations $D_+ \Phi \Db_+ \Phib$, or required dependence on other 
 dimensionless variables like $\lambda (D_+ D_- \Phi)^2$, but these complications do not arise in the case where the undeformed 
 theory only has a K\"ahler potential.

On dimensional grounds, $f$ must be proportional to $\lambda$ times a function of the dimensionless combinations $\lambda x$ 
and $\lambda y$. Thus, although the differential equation for $f$ determined by (\ref{bg_def_unsimp}) is complicated, 
one can solve order-by-order in $\lambda$. The solution to $\mathcal{O} ( \lambda^3 )$ is
\begin{align}
    f ( \lambda, x, \xb, y ) =&\, \frac{\lambda}{2} + \lambda^2 \left( \frac{x + \xb}{4} - \frac{3}{4} y \right)
     \cr &
      + \lambda^3 \left( \frac{x^2 + \xb^2 + 3 x \xb}{8} + \frac{37}{24} y^2 - \frac{25}{24} \left( x + \xb \right) y \right) + \cdots~ .
\label{series-1}\end{align}
We were unable to find a closed-form expression for $f$ to all orders in $\lambda$. However, the differential equation 
simplifies dramatically when we impose the equations of motion for the theory, and in this case one can write down an exact formula.
This is similar to the $T\Tb$ flow of the free action for a real $\cN=(1,1)$ scalar multiplet that was analyzed in
\cite{Baggio:2018rpv,Chang:2018dge}.

We claim that, on-shell, one may drop any terms where $y \sim ( D_+ D_- \Phi ) ( \Db_+ \Db_- \Phib )$ multiplies the
 four-fermion term $|D\Phi|^4 \equiv D_+ \Phi \Db_+ \Phib D_- \Phi \Db_- \Phib$. This is shown explicitly in Appendix 
 \ref{appendix:on-shell} and follows directly from the superspace equation of motion and nilpotency of the fermionic terms
  $D_{\pm} \Phi$ and $\Db_{\pm} \Phib$. It is also an intuitive statement associated to the 
fact that for these models, on-shell, $\cN=(2,2)$
supersymmetry is not broken.
In fact, note that the superfields $( D_+ D_- \Phi )$ and
$( \Db_+ \Db_- \Phib )$ have as their lowest components the 
auxiliary fields $F$ and $\overline{F}$. If supersymmetry is not broken, the vev of $F$ has to be zero, 
$\langle F\rangle = 0$, which implies that
the auxiliary field $F$ is on-shell at least quadratic 
in fermions and, more precisely, can be proven
to be at least linear in $\psi_\pm=D_\pm\Phi|_{\q=0}$
and $\bar{\psi}_\pm=\Db_\pm\overline{\Phi}|_{\q=0}$.
From this argument it follows that on-shell 
$( D_+ D_- \Phi )$ is at least linear in 
$D_\pm\Phi$ and $\Db_\pm\overline{\Phi}$, and then the two conditions
$( D_+ D_- \Phi )|D\Phi|^4=0$ and $y|D\Phi|^4=0$ follow.

After removing  from \eqref{bg_def_unsimp}
the $y$-dependent terms which vanish on-shell,
we find a simpler differential equation 
for the function $f$,
\begin{align}
    \frac{df}{d \lambda} &= \frac{1}{2} \left\{ - x \xb \left[ f + x \frac{\partial f}{\partial x} + \xb \frac{\partial f}{\partial \xb} \right]^2 
    + \left[ 1 + (x + \xb) f + x \xb \left( \frac{\partial f}{\partial x} + \frac{\partial f}{\partial \xb} \right) \right]^2 \right\} ~,
\end{align}
whose solution is
\begin{align}
    f( \lambda , x, \xb, y=0) = \frac{\lambda}{1 - \frac{\lambda}{2} ( x + \xb )  + \sqrt{1 - \lambda ( x + \xb )  
    + \frac{\lambda^2}{4} \left( x - \xb \right)^2 }}
    ~.
\end{align}
Thus we have shown that the supercurrent-squared deformed Lagrangian at finite $\lambda$ is equivalent on-shell 
to the following superspace Lagrangian
\begin{align}
    \mathcal{L}_{\lambda} &= \int d^4 \theta \, \left( K ( \Phi, \Phib ) 
    + \frac{\lambda K_{\Phi \Phib}^2 D_+ \Phi \Db_+ \Phib D_- \Phi \Db_- \Phib }{1 - \frac{1}{2} \lambda K_{\Phi \Phib}^2 A 
    + \sqrt{1 - \lambda K_{\Phi \Phib}^2 A  + \frac{1}{4} \lambda^2 K_{\Phi \Phib}^4 B^2 } } \right)~ ,
    \label{bg_on_shell}
\end{align}
where
\begin{align}
    A = \partial_{++} \Phi \partial_{--} \Phib + \partial_{++} \Phib \partial_{--} \Phi ~,\qquad
    B = \partial_{++} \Phi \partial_{--} \Phib - \partial_{++} \Phib \partial_{--} \Phi ~. 
\end{align}

When $K ( \Phi , \Phib ) = \Phib \Phi$, 
it is simple to show that this model represents 
an  $\mathcal{N} = (  2, 2 )$ 
off-shell supersymmetric extension of the $D=4$ Nambu-Goto string in an appropriate gauge---
often referred to as a static gauge in presence of a $B$ field, though it can be more naturally described as uniform light-cone 
gauge~\cite{Arutyunov:2004yx,Arutyunov:2005hd} (see refs.~\cite{Baggio:2018gct,Frolov:2019nrr} for a discussion of this point).
In particular, by setting various component fields to zero and performing the superspace integrals, one can show that 
(\ref{bg_on_shell}) matches the expected answer for $T \Tb$ deformations in previously known non-supersymmetric cases. 
For instance, setting the fermions to zero and integrating out the auxiliary fields 
$F$ and $\overline{F}$
gives  the $T \Tb$ deformation of the complex free boson
$\phi$, whose Lagrangian is
\begin{equation}
\label{eq:bosonTTbar}
 {\cal L}_{\lambda,\text{bos}}= 
 \frac{\sqrt{1+2 \lambda  a +\lambda^2 b^2}-1}{4\lambda}
 =
  \frac{  a}{4 } - \lambda \frac{ \pa_{++} \phi \pa_{--}\phi \pa_{++} \bar \phi \pa_{--} \bar\phi }
 {  1+\lambda a+\sqrt{1+2 \lambda  a +\lambda^2 b^2}  } 
 ~,
\end{equation}
where
\begin{equation}
  \label{eq:xydef}
 a=\pa_{++} \phi \pa_{--} \bar \phi +\pa_{++}  \bar \phi \pa_{--}  \phi
 ~, \qquad 
b= \pa_{++} \phi \pa_{--} \bar \phi -\pa_{++}  \bar \phi \pa_{--}  \phi ~.
\end{equation}
The Lagrangian \eqref{eq:bosonTTbar} indeed describes the $D=4$ light-cone gauge-fixed Nambu-Goto string model.

Alternatively, setting all the bosons to zero
in \eqref{bg_on_shell} can be shown to give the $T \Tb$ deformation of a complex free fermion. 
These calculations are similar to those in the case of the $(0,2)$ supercurrent-squared action, 
which are presented in \cite{Jiang:2019hux}. In fact, it can even be easily shown
that an $\cN=(0,2)$ truncation of \eqref{bg_on_shell}
gives precisely the $T\Tb$ deformation of a free $\cN=(0,2)$ chiral multiplet that was derived in \cite{Jiang:2019hux}.

It is worth highlighting that, unlike the $\cN=(2,2)$ case,
an off-shell 
$(0,2)$ chiral scalar multiplet contains only physical degrees
of freedom and no auxiliary fields.
Interestingly, related to this fact, it turns out that
(up to integration by parts and total derivatives)
the $\cN=(0,2)$ off-shell supersymmetric extension of the  
$D=4$ Nambu-Goto string action in light-cone gauge is unique
and precisely matches the off-shell $T\Tb$ deformation of a free 
$\cN=(0,2)$ chiral multiplet action \cite{Jiang:2019hux}.

In the $\cN=(2,2)$ case, because of the presence of the auxiliary 
field $F$ in the chiral multiplet $\Phi$, 
there are an infinite set of inequivalent $\cN=(2,2)$ off-shell extensions
of the Lagrangian \eqref{eq:bosonTTbar} that are all equivalent 
on-shell. 
A representative of these equivalent actions is described by
(\ref{bg_on_shell}) when $K(\F,\overline{\F})=\overline{\F}\F$.

The non-uniqueness of dynamical systems described by actions
of the form (\ref{bg_on_shell})
can also be understood by noticing that, for example,
it is possible to perform a class of redefinitions that leaves the action (\ref{bg_on_shell}) invariant on-shell. 
As a (very particular)  example, 
note that we are free to perform a shift 
of the form
\begin{align}
    D_+ \Db_- \left( \Db_+ \Phib D_- \Phi \right) \longrightarrow D_+ \Db_- \left( \Db_+ \Phib D_- \Phi \right)
     + a \left( D_+ D_- \Phi + \Db_+ \Db_- \Phib \right)^2
    \label{field_redef}
\end{align}
for any real number $a$. In terms of $A$ and $B$, (\ref{field_redef}) implements the shifts
\begin{align}\begin{split}
    A \longrightarrow A + a \left( \left( D_+ D_- \Phi \right)^2 + 2 y + \left( \Db_+ \Db_- \Phib \right)^2 \right)~,\qquad
    B \longrightarrow B 
        \label{field_redef-2}
\end{split}\end{align}
in (\ref{bg_on_shell}).
The resulting Lagrangian would enjoy the same on-shell 
simplifications described in  Appendix \ref{appendix:on-shell}
and would turn out to be on-shell equivalent to
the Lagrangian (\ref{bg_on_shell}).
In this infinite set of on-shell
equivalent actions, a particular choice would represent
an exact solution of the $T\Tb$ flow equation
\eqref{sc2_kahler_flow}--\eqref{bg_def_unsimp},
whose leading terms in a $\lambda$ series expansion are given in \eqref{series-1}.
Another representative in this on-shell equivalence class
is the simplified model described by (\ref{bg_on_shell}).

These types of redefinition 
and on-shell equivalentness 
are not a surprise, nor really new. 
In fact, they are  of the same nature as redefinitions 
that have  been studied in detail in \cite{GonzalezRey:1998kh}
(see also \cite{Kuzenko:2011tj} for a description of these 
 types of ``trivial symmetries'')
 in the context of 
$D=4$ $\cN=1$ chiral and linear superfield models possessing a non-linearly realised additional supersymmetry
\cite{Bagger:1997pi,GonzalezRey:1998kh}. 
 As in (\ref{field_redef-2}), the field redefinition in 
this context does not affect the dynamics of the physical fields---
it basically corresponds only to an arbitrariness in the definition of the 
auxiliary fields that always appear quadratically
in the action and then are set to zero (up to fermion terms
that will not contribute due to nilpotency in the action)
on-shell. 
Although here we only focused on discussing the on-shell ambiguity
of the solution of the $\cN=(2,2)$ $T\Tb$ flow,
we expect that the exact solution of the flow equations
with $y$ nonzero \eqref{sc2_kahler_flow}--\eqref{bg_def_unsimp} 
can be found by a 
field redefinition of the kind we made in the action (\ref{bg_on_shell}). 

It is also interesting to note that similar
freedoms and field redefinitions 
are also described in the construction
of $D=4$ $\cN=1$ supersymmetric Born-Infeld actions;  see, for example,~\cite{Bagger:1996wp}.
In fact, as will be analyzed in more detail elsewhere 
\cite{SUSYDBITTbar},
it can be shown that the Lagrangian \eqref{bg_on_shell}
is structurally of the type described by Bagger and Galperin
for the $D=4$ $\cN=1$ supersymmetric Born-Infeld action 
\cite{Bagger:1996wp}. 
The equivalence can be formally shown by identifying 
$W_+ = \Db_+ \Phib$, $W_- = D_- \Phi$, $W^2 = \Db_+ \Phib D_- \Phi$, and 
$D^\alpha W_\alpha = D_+ D_- \Phi + \Db_- \Db_+ \Phib$ to match their conventions. 
As a consequence, we can show that our solution for the $T\Tb$ flow possesses a second non-linearly realised $\cN=(2,2)$ supersymmetry, besides the $(2,2)$ supersymmetry which is made manifest by the superspace construction. This property is analyzed in detail in \cite{SUSYDBITTbar}. We note that the presence of a second supersymmetry is analogous to what happens in the $\cN=(0,2)$ case~\cite{Jiang:2019hux}.

\subsection{Adding a superpotential}

Now suppose we begin with an undeformed theory that has a superpotential $W(\Phi)$,
\begin{align}
    \mathcal{L}^{(0)} = \int d^4 \theta \, K ( \Phi , \Phib ) + \left( \int d^2 \theta \, W ( \Phi ) \right) 
    + \left( \int d^2 \thetab \, \overline{W} ( \Phib ) \right)~ .
    \label{lag_with_W}
\end{align}
As shown in Appendix \ref{appendix:supercurrent_calculation}, the superpotential F-term gives a contribution 
$\delta \mathcal{V} = 2 W(\Phi)$ to the field $\mathcal{V}$ which appears in supercurrent-squared. 
To leading order in the deformation parameter, the Lagrangian takes the form 
\begin{align}
    \mathcal{L}^{(0)} &\to \mathcal{L}^{(0)} + \mathcal{L}^{(1)} \nonumber \\
    &= \mathcal{L}^{(0)} 
    + \lambda \int d^4 \theta \, 
    \left( \frac{1}{2} K_{\Phi \Phib}^2 D_+ \Phi \Db_+ \Phib D_- \Phi \Db_- \Phib + W(\Phi) \Wb ( \Phib ) \right) ~.
\end{align}
In addition to the four-fermion term which we saw in section \ref{subsec:kahler}, we see that the deformation modifies the 
K\"ahler potential, adding a term proportional to $|W ( \Phi )|^2$.

Next consider the second order term in $\lambda$. For convenience, we use the combination 
$|D\Phi|^4 = D_+ \Phi \Db_+ \Phib D_- \Phi \Db_- \Phib$, which is the four-fermion combination that appeared at first order. Then
\begin{align}
    \mathcal{L}^{(2)} 
    =
    \frac{\lambda^2}{4}
    \int d^4 \theta \,
     \Big( x + \xb - 3 y 
     -2 | W' ( \Phi ) |^2 
     +W D_- D_+ +\Wb \Db_+ \Db_-\Big) |D\Phi|^4
    ~. 
    \label{superpotential_second_order}
\end{align}
The new terms involving supercovariant derivatives of $|D\Phi|^4$ will generate contributions with two fermions in the D-term.

As we continue perturbing to higher orders, the form of the superspace Lagrangian becomes more complicated. 
It is no longer true that the supercurrent-squared flow closes on a simple ansatz with one undetermined function, 
as it did in the case with only a K\"ahler potential. Indeed, the finite-$\lambda$ deformed superspace Lagrangian
 in the case with a superpotential will depend not only on the variables $x$, $\xb$, 
 and $y$ as in section \ref{subsec:kahler}, 
 but also, for example,
 on combinations like $\partial_{++} \Phi \Db_+ \Db_- \Phib$, which can appear multiplying the two-fermion term
  $D_- \Phi \Db_- \Phib$ in the superspace Lagrangian. To find the full solution, one would need to determine several functions contributing to the D-term---one multiplying the four-fermion term $|D\Phi|^4$ as in the K\"ahler case; one for the deformed K\"ahler potential which may 
  now depend on $x$, $y$, and other combinations; and four functions multiplying the two-fermion terms 
  $D_+ \Phi D_- \Phi$, $D_+ \Phi \Db_- \Phib$, etc. Each function can depend on several dimensionless combinations. 

In the presence of a superpotential, 
the situation might further be complicated by the fact that 
supersymmetry can be spontaneously broken.
This would make it impossible, for example, to use on-shell
simplifications like $y|D\Phi|^4=0$ that we employed in the section~\ref{subsec:kahler}, where 
supersymmetry is never spontaneously broken.

It should be clear that the case with a superpotential is significantly
more involved and rich than just a pure K\"ahler potential.
In this case, we have not attempted to find a solution of the 
$T\Tb$ flow equation in closed form.
However, it is evident from the form of supercurrent-squared 
eq.~\eqref{sc2_kahler_flow}---which is always written as a D-term integral of current bilinears---
that this deformation will only affect the $D$ term and not the $\cN=(2,2)$ superpotential $W$ appearing in the chiral integral. 
Therefore the superpotential, besides being protected from perturbative quantum corrections, is also protected from corrections
 along the supercurrent-squared flow.

\subsection{The physical classical potential} 

In view of the difficulty of finding the all-orders deformed superspace action for a theory with a superpotential, 
we now consider the simpler problem of finding the local-potential approximation (or zero-momentum potential) for the bosonic complex scalar 
$\phi$ contained in the superfield $\Phi$. 
We stress that our analysis here is purely classical
and we will make a couple of  
comments about possible quantum effects 
later in this section.
For simplicity, we will also restrict to 
the case in which the K\"ahler 
potential is flat, $K(\Phi,\overline{\Phi})=\overline{\Phi}\F$.
By ``zero-momentum potential'' we mean the physical potential
$V ( \phi )$ which appears in the  
Lagrangian after performing the superspace integral in the deformed theory and then setting $\partial_{\pm \pm} \phi = 0$. 
For instance, consider the undeformed Lagrangian
\begin{align}
    \mathcal{L}^{(0)} = \int d^4 \theta \, \Phib \Phi + \int d^2 \theta \, W ( \Phi ) + \int d^2 \thetab \, \Wb ( \Phib ) ~.
\end{align}
When we ignore all terms involving derivatives and the fermions $\psi_{\pm}$, the only contributions to the physical Lagrangian 
(after performing the superspace integral) come from an $|F|^2$ term from the kinetic term, plus the term 
$W(\Phi) = W ( \phi ) + W'(\phi) \theta^+ \theta^- F$. This gives us the zero-momentum, zero-fermion component action
\begin{align}
    S = \int d^2 x \, \left(|F|^2 + W' ( \phi ) F + \Wb' ( \phib) \Fb \right)~ .
    \label{superpotential_integrated}
\end{align}
We may integrate out the auxiliary field $F$ using its equation of motion $\Fb = - W' ( \phi )$, which yields
\begin{align}
    S = \int d^2 x \, \left( - | W' ( \phi ) |^2 \right)~ ,
\end{align}
so the zero-momentum potential for $\phi$ is $V = | W' ( \phi )|^2$, as expected.
Note that the previous potential might have extrema that breaks $\cN=(2,2)$ supersymmetry
while supersymmetric vacua will always set 
$\langle F\rangle=\langle W'(\phi)\rangle = 0$. We will
assume supersymmetry of the undeformed theory 
not to be spontaneously broken in our discussion.

Now suppose we deform by the supercurrent-squared operator to second order in $\lambda$, 
which gives the superspace expression (\ref{superpotential_second_order}). If we again perform the superspace 
integral and discard any terms involving derivatives or fermions, we now find the physical Lagrangian
\bea
\mathcal{L} \big\vert_{\partial_{\pm \pm} \phi = 0 } 
&=& |F|^2 + F W' + \Fb \Wb' + \lambda \left( \frac{1}{2} |F|^4 - |F|^2 |W'|^2 \right) 
+ \frac{1}{4} \lambda^2 | F |^4 \left( W' F + \Wb' \Fb \right) 
\non\\
    && - \frac{1}{2} \lambda^2 | W' |^2 | F |^4 + \frac{3}{4} \lambda^2 | F |^6 ~.
    \label{second_order_W_action}
\eea
Remarkably, the equations of motion for the auxiliary $F$
in (\ref{second_order_W_action}) admit the solution 
$F = - \Wb' ( \phib ) $, $\Fb = - W' ( \phi )$, which is the same 
as the unperturbed solution.
This for instance implies that if we  start
from a supersymmetric vacua in the undeformed theory we will 
remain supersymmetric along the $T\Tb$ flow.
On the one hand, this
is not a surprise considering that we know the $T\Tb$
flow preserves the structure of the spectrum, and in particular should leave a zero-energy supersymmetric vacuum unperturbed.
On the other hand, it is a reassuring check to see 
this property  explicitly appearing in our analysis.

Returning to \eqref{second_order_W_action} and
integrating out the auxiliary fields gives
\begin{align}
    \mathcal{L} \big\vert_{\partial_{\pm \pm} \phi = 0 } = - |W'(\phi)|^2 - \frac{1}{2} \lambda | W' ( \phi ) |^4 
    - \frac{1}{4} \lambda^2 | W' ( \phi ) |^6~ .
    \label{leading_geometric}
\end{align}
These are the leading terms in the geometric series $\frac{-|W'|^2}{1 - \frac{1}{2} \lambda |W'|^2}$.
 In fact, up to conventions for the scaling of $\lambda$, one could have predicted this outcome from the form of the
  supercurrent-squared operator and the known results for $T \Tb$ deformations of a bosonic theory with a potential 
  \cite{Cavaglia:2016oda}.
We know that, up to terms which vanish on-shell, the effect of adding supercurrent-squared to the physical Lagrangian
 is to deform by the usual $T \Tb$ operator. However, in the zero-momentum sector, we see that the $T \Tb$ deformation 
 reduces to deforming by the square of the potential:
\begin{align}
    T \Tb \Big\vert_{\partial_{\pm \pm} \phi = 0} = \mathcal{L}^2 \Big\vert_{\partial_{\pm \pm} \phi = 0} = V^2 ~.
\end{align}
Therefore, it is easy to solve for the deformed potential if we deform a physical Lagrangian
 $\mathcal{L} = f ( \lambda, \partial_{\pm \pm} \phi ) + V ( \lambda , \phi )$ by $T \Tb$, 
 since the flow equation for the potential term is simply
\begin{align}
    \partial_\lambda \mathcal{L} = \frac{\partial V}{\partial \lambda} = V^2~ ,
\end{align}
which admits the solution
\begin{align}
    V ( \lambda , \phi ) = \frac{V ( 0, \phi ) }{1 - \lambda V ( 0 , \phi ) } ~.
\end{align}
We can apply this result to the Lagrangian (\ref{superpotential_integrated}), treating the entire expression involving
 the auxiliary field $F$ as a potential (since it is independent of derivatives). The deformed theory has a zero-momentum
  piece which is therefore equivalent to
\begin{align}
    S (\lambda ) \Big\vert_{\partial_{\pm \pm} \phi = 0} = \int d^2 x \, \frac{\left(|F|^2 + W' ( \phi ) F 
    + \Wb' ( \phib) \Fb \right)}{1 - \lambda \left(|F|^2 + W' ( \phi ) F + \Wb' ( \phib) \Fb \right)}~ ,
\end{align}
at least on-shell. Integrating out the auxiliary now gives
\begin{align}
    S (\lambda ) \Big\vert_{\partial_{\pm \pm} \phi = 0} = \int d^2 x \,  \frac{- |W'(\phi)|^2 }{1 - \lambda |W'(\phi)|^2 }
    \label{deformed_superpotential}
\end{align}
as the deformed physical potential. This matches the first few terms of (\ref{leading_geometric}), 
up to a convention-dependent factor of $\frac{1}{2}$ in the scaling of $\lambda$.

Now one might ask what superspace Lagrangian would yield the physical action (\ref{deformed_superpotential}) 
after performing the $d \theta$ integrals. One candidate is
\begin{align}
    \mathcal{L} (\lambda ) \Big\vert_{\partial_{\pm \pm} \phi = 0} \sim \int d^4 \theta \, 
    \left( \Phib \Phi - \lambda | W (\Phi) |^2 \right) + \int d^2 \theta \, W ( \Phi ) + \int d^2 \thetab \, \Wb ( \Phib ) ~,
    \label{deformed_superpotential_superspace}
\end{align}
where here $\sim$ means ``this superspace Lagrangian gives an equivalent zero-momentum physical potential for the boson 
$\phi$ on-shell.''

It is important to note that (\ref{deformed_superpotential_superspace}) is \emph{not} the true solution for the deformed superspace 
Lagrangian using supercurrent-squared. The genuine solution involves a four-fermion term, all possible two-fermion terms, and more 
complicated dependence on the variable $y = \lambda ( D_+ D_- \Phi ) ( \Db_+ \Db_- \Phib )$ in the zero-fermion term. However, if 
one were to perform the superspace integral in the true solution and then integrate out the auxiliary field $F$ using its equation of 
motion, one would obtain the same zero-momentum potential for $\phi$ as we find by performing the superspace integral in 
(\ref{deformed_superpotential_superspace}) and integrating out $F$. 

The form (\ref{deformed_superpotential_superspace}) is interesting because it shows that the effect of supercurrent-squared on the 
physical potential for $\phi$ can be interpreted as a change in the K\"ahler metric, which for this Lagrangian is
\begin{align}
    K_{\Phi \Phib} = 1 - \lambda | W' ( \Phi ) |^2~ .
    \label{effective_kahler}
\end{align}
When one performs the superspace integrals in (\ref{deformed_superpotential_superspace}), the result is
\begin{align}
    \mathcal{L} \Big\vert_{\partial_{\pm \pm} \phi = 0} = K_{\Phi \Phib} | F |^2 + W' ( \phi ) F + \Wb' ( \phib ) \Fb ~, 
\end{align}
which admits the solution $F = - \frac{\Wb'(\phib)}{K_{\Phi \Phib}}$. Substituting this solution gives
\begin{align}
    \mathcal{L} \Big\vert_{\partial_{\pm \pm} \phi = 0} = \frac{ - | W' ( \phi ) |^2 }{K_{\Phi \Phib}} 
    = \frac{- | W' ( \phi )|^2 }{1 - \lambda | W' ( \phi ) |^2}~ ,
\end{align}
which agrees with (\ref{deformed_superpotential}).

As already mentioned, supersymmetric vacua of the original, undeformed, theory are associated with critical points of the 
superpotential $W(\phi)$. Any vacuum of the undeformed theory will persist in the deformed theory: near a point where 
$W' ( \phi ) = 0$, we see that the physical potential $V ( \phi ) = \frac{ | W' |^2}{1 - \lambda |W'|^2}$ also vanishes (away from the pole 
$|W'|^2 = \frac{1}{\lambda}$, the deformed potential is a monotonically increasing function of $|W'|^2$). Further, the auxiliary field $F$ 
does not acquire a vacuum expectation value because $F = - \Wb' ( \phib )$ remains a solution to its equations of motion in the 
deformed theory. Once more, this indicates that supersymmetry is unbroken along the whole $T\Tb$ flow if is in the undeformed
theory.

However, this classical analysis suggests that the soliton spectrum of the theory has changed dramatically at any finite deformation 
parameter $\lambda$. There are now generically poles in the physical potential $V(\phi)$ at points where $|W'|^2 = \frac{1}{\lambda}$ 
which might separate distinct supersymmetric vacua of the theory. 
For instance, if the original theory had a 
double-well superpotential 
with two critical points $\phi_1$, $\phi_2$ where 
$W'(\phi_i) = 0$, then this undeformed theory supports BPS 
soliton solutions which 
interpolate between these two vacua. But if the superpotential 
$W$ reaches a value of order $\frac{1}{\lambda}$ at some point 
between $\phi_1$ and $\phi_2$, then this soliton solution appears 
naively forbidden in the deformed theory because it requires 
crossing an infinite potential barrier. Another way of seeing 
this is by considering the effective K\"ahler potential (\ref{effective_kahler}), 
which would change sign at some point between the two 
supersymmetric vacua in the deformed theory and thus give rise 
to a negative-definite K\"ahler metric.

Our discussion has been purely classical. As we emphasised in the introduction, a fully quantum analysis of this problem is desirable, though subtle because of the non-local nature of the $T\Tb$ deformation. The advantage of performing such an analysis in models with extended supersymmetry is that holomorphy and associated non-renormalization theorems provide control over the form of any possible quantum corrections. For example, the superpotential for the models studied in this work is not renormalized perturbatively along the flow. It would be interesting to examine the structure of perturbative quantum corrections along the lines of~\cite{Rosenhaus:2019utc}, but in superspace with manifest supersymmetry. It should be possible to absorb any quantum corrections visible in perturbation theory by a change in the $D$-term K\"ahler potential
meaning that at least the structure  of the
supersymmetric vacua would be preserved.


\section*{Acknowledgements}

We thank Marco Baggio for discussions and collaboration in the initial stages of this project.
C.~F. acknowledges support from the University of Chicago's divisional MS-PSD program. C.~C., C.~F. and S.~S. are supported in 
part by NSF Grant No. PHY1720480. 
 This work is partially supported through a research grant of the Swiss National Science Foundation, 
 as well as by the NCCR SwissMAP, funded by the Swiss National Science Foundation.
 The work of A.~S.\ was also supported by ETH Career Seed Grant SEED-23 19-1.
The work of G.~T.-M. is supported by the Albert Einstein Center for Fundamental Physics, University of Bern,
and by the Australian Research Council (ARC) Future Fellowship FT180100353.
We would also like to thank the organizers and participants of the workshop on \emph{``$T \Tb$ and Other Solvable Deformations of 
Quantum Field Theories''} 
for providing a stimulating 
atmosphere,
 and the Simons Center for Geometry and Physics for hospitality and partial support 
during the final stages of this work. We thank the participants of the workshop ``New frontiers of integrable deformations'' at Villa Garbald in Castasegna for stimulating discussions.

\appendix


\section{The \texorpdfstring{$\cS$}{S}-multiplet in components}
\label{components-S}

In this Appendix we provide
the component expansion
of the superfields of the $\cS$-multiplet introduced in section \ref{section-S-multiplet}.
The results presented below are equivalent to the results
first obtained in \cite{Dumitrescu:2011iu} up to 
differences in notation.

The constraints~\eqref{conservation-S} 
are solved in terms of component fields by, 
\bea
\label{comp-S}
{\cal S}_{\pm\pm}  
&= &
 j_{\pm\pm} - i \q^\pm S_{\pm\pm\pm} 
 - i \q^\mp \left( S_{\mp\pm\pm} \mp 2 \sqrt 2 i \bar \rho_\pm\right) 
 - i \qb^\pm \bar S_{\pm\pm\pm} \cr
&& 
- i \qb^\mp \left(\bar S_{\mp\pm\pm} \pm 2 \sqrt2 i \rho_\pm\right) 
- \q^\pm \qb^\pm T_{\pm\pm\pm\pm} 
+ \q^\mp \qb^\mp \left(A \mp { \frac{k + k'}{2}}\right) \cr
&& 
+ i \q^+ \q^- \bar Y_{\pm\pm} + i \qb^+ \qb^- Y_{\pm\pm} \pm i \q^+ \qb^- \bar G_{\pm\pm} \mp i \q^- \qb^+ G_{\pm\pm} 
\cr
&& 
\mp \hf \q^+ \q^- \qb^\pm \pa_{\pm\pm} S_{\mp\pm\pm} \mp \hf \q^+ \q^- \qb^\mp \pa_{\pm\pm}
\left(S_{\pm\mp\mp} \pm 2 \sqrt 2 i \bar \rho_\mp\right)\cr
&&
 \mp \hf \qb^+ \qb^- \q^\pm \pa_{\pm\pm} \bar S_{\mp\pm\pm} \mp \hf \qb^+ \qb^-\q^\mp \pa_{\pm\pm} 
 \left( \bar S_{\pm\mp\mp} \mp 2 \sqrt 2  i \rho_\mp\right)\cr
&& + {\frac{1}{4}} \q^+ \q^- \qb^+ \qb^- \pa_{\pm\pm}^2 j_{\mp\mp}~.
\eea
Let us introduce the usual useful combinations: 
$y^{\pm\pm} = x^{\pm\pm} - \frac{i}{2}  \q^\pm\qb^\pm$ and $\tilde y^{\pm\pm} = x^{\pm\pm} \mp \frac{i}{2} \q^\pm \qb^\pm$. 
The chiral superfields~$\chi_\pm$ are
\bsubeq
\label{comp-chi}
\bea
\chi_+ &=&
 - i \lambda_+(y) - i \q^+ \bar G_{++}(y) +  \q^- \left(E(y) + {\frac{k}{2}}\right) + \qb^- C^{(-)}  + \q^+ \q^- \pa_{++} \bar \lambda_-(y)~,
~~~~~~\\
\chi_- &=&
 - i \lambda_-(y) - \q^+ \left(\bar E(y) - {\frac{k}{2}}\right) + i \q^- G_{--}(y) - \qb^+ C^{(+)}- \q^+ \q^- \pa_{--} \bar \lambda_+(y)~,\\
\lambda_\pm &=& \pm\bar S_{\mp\pm\pm} + \sqrt 2 i \rho_\pm~,
\\
E &=& \hf \left( \Theta - A \right)+ {\frac{i}{4}} \left( \pa_{++} j_{--} - \pa_{--} j_{++}\right)~,\\
0&=& \pa_{++} G_{--} - \pa_{--} G_{++} ~,
\eea
\esubeq
and the twisted-(anti-)chiral superfields~$\cY_\pm$ are given by
\bsubeq
\label{comp-Y}
\bea
\cY_+ &=& \sqrt 2 \rho_+ (\bar {\tilde y}) +  \q^- \left( F(\bar {\tilde y}) 
+ {\frac{k'}{2}}\right) - i \qb^+ Y_{++}(\bar {\tilde y}) - \qb^- C^{(-)}+ \sqrt2 i \q^- \bar \q^+ \pa_{++} \rho_-(\bar {\tilde y})
~,~~~~~~~~~
\\
\cY_- &=& \sqrt 2 \rho_-(\tilde y) -  \q^+ \left( F(\tilde y) - {\frac{k'}{2}}\right) + \qb^+ C^{(+)}
- i \qb^- Y_{--}(\tilde y) + \sqrt 2 i \q^+ \qb^- \pa_{--} \rho_+ (\tilde y)~,\\
F &=& - \hf \left( \Theta + A\right) - {\frac{i}{4}} \left(\pa_{++} j_{--} + \pa_{--} j_{++}\right)~,\\
0&=&\pa_{++} Y_{--} - \pa_{--} Y_{++} ~.
\eea
\esubeq

For the FZ-multiplet defined by the constraints \eqref{FZ-2}, the $\cS$-multiplet reduces to a set of $4+4$ real independent
component fields described by the $j_{\pm\pm}$ $U(1)_A$ axial conserved $R$-symmetry current  
($\pa_{++} j_{--} - \pa_{--} j_{++}=0$). In addition, there is a complex scalar field $v(x)$,
see eq.~\eqref{FZ-3},
together with the independent supersymmetry current and energy momentum tensor:
\bsubeq
\bea
 S_{\pm\pm\pm}(x)
&:=&
 ~~i  D_{\pm}\cJ_{\pm\pm}(\z)|_{\q=0}
~,\\
\bar{S}_{\pm\pm\pm}(x)
&:=&
- i  \Db_{\pm}\cJ_{\pm\pm}(\z)|_{\q=0}
~,\\
 S_{\mp\pm\pm}(x)
&:=&
- i  D_{\mp}\cJ_{\pm\pm}(\z)|_{\q=0}
=
\pm i  \Db_{\pm}\overline{\cV}(\z)|_{\q=0}
~,
\\
\bar{S}_{\mp\pm\pm}(x)
&:=&
~~ i  \Db_{\mp}\cJ_{\pm\pm}(\z)|_{\q=0}
=\mp i  D_{\pm}\cV(\z)|_{\q=0}
~,
\\
T_{\pm\pm\pm\pm}(x)
&:=&
~~\hf{[}D_\pm,\Db_\pm{]}\cJ_{\pm\pm}(\z)|_{\q=0}
~,
\\
\Theta(x)
&:=&
-\hf{[}D_+,\Db_+{]}\cJ_{--}(\z)|_{\q=0}
=-\hf{[}D_-,\Db_-{]}\cJ_{++}(\z)|_{\q=0}
\non\\
&=&-\hf D_+D_-\cV(\z)|_{\q=0}
+\hf \Db_+\Db_-\overline{\cV}(\z)|_{\q=0}
~.
\eea
\esubeq
For the FZ-multiplet, the following relation holds:
\bea
{\cal J}_{\pm\pm}  
&= & j_{\pm\pm} 
- i \q^\pm S_{\pm\pm\pm} 
 - i \qb^\pm \bar S_{\pm\pm\pm} 
+ i \q^\mp S_{\mp\pm\pm}
+ i \qb^\mp \bar S_{\mp\pm\pm} 
\cr
&&
- \q^\pm \qb^\pm T_{\pm\pm\pm\pm} + \q^\mp \qb^\mp \Theta
+ i   \q^+ \q^- \pa_{\pm\pm}\bar v + i  \qb^+ \qb^- \pa_{\pm\pm} v
\cr
&& \mp \hf \q^+ \q^- \qb^\pm \pa_{\pm\pm} S_{\mp\pm\pm} 
\pm \hf \q^+ \q^- \qb^\mp \pa_{\pm\pm}S_{\pm\mp\mp}
\cr
&&
 \mp \hf \qb^+ \qb^- \q^\pm \pa_{\pm\pm} \bar S_{\mp\pm\pm} 
 \pm \hf \qb^+ \qb^-\q^\mp \pa_{\pm\pm}  \bar S_{\pm\mp\mp} 
\cr
&&
 + {\frac{1}{4}} \q^+ \q^- \qb^+ \qb^- \pa_{\pm\pm}^2 j_{\mp\mp}
 ~.
\eea
Moreover, the chiral superfields~$\chi_\pm$ are set to zero 
and the twisted-(anti-)chiral superfields~$\cY_\pm=D_\pm \cV$ are given by
\bsubeq
\bea
\cY_+ &=&   i\bar S_{-++} (\bar {\tilde y}) +  \q^- \, 
G(\bar {\tilde y}) 
- i \qb^+ \pa_{++}v(\bar {\tilde y}) 
+ \q^- \bar \q^+ \pa_{++}  \bar S_{+--}(\bar {\tilde y})~,
\\
\cY_- &=& - i \bar S_{+--}(\tilde y) -  \q^+ G(\tilde y) 
- i \qb^- \pa_{--}v(\tilde y) 
+ \q^+ \qb^- \pa_{++}  \bar S_{---}(\tilde y)~,
\\
G &=& - \Theta - {\frac{i}{2}} \pa_{++} j_{--}
~.
\eea
\esubeq

\section{Details of the supercurrent multiplet calculation}\label{appendix:supercurrent_calculation}

In this Appendix, we compute the fields $\mathcal{J}_{\pm \pm}$ and $\sigma$ appearing in
 the FZ-multiplet for Lagrangians of a chiral superfield $\Phi$ with the general form
\begin{align}
    \mathcal{L}_0 = \left( \int d^4 \theta \, \mathcal{A} ( \Phi, D_{\pm} \Phi , D_+ D_- \Phi, \partial_{\pm \pm} \Phi , \text{c.c.} ) \right) 
    + \left( \int d^2 \theta \, W ( \Phi ) \right) + \left( \int d^2 \thetab \, \Wb ( \Phib ) \right) ~, 
    \label{general_susy_lag}
\end{align}
where ``\text{c.c.}'' indicates dependence on the conjugates $\Phib, \Db_{\pm} \Phib, \Db_+ \Db_- \Phib$, and 
$\partial_{\pm \pm} \Phib$. To do this, we will minimally couple the theory to supergravity using the old-minimal supergravity 
formulation and extract the currents which couple to the metric superfield $H^{\pm \pm}$ and the chiral compensator $\sigma$. 
The minimal coupling prescription involves promoting $\mathcal{L}_0$ to%
\footnote{Conforming to 
notation of \cite{Grisaru:1994dm,Grisaru:1995dr,Grisaru:1995kn,Grisaru:1995py,Gates:1995du}, 
in this section we will sometimes 
use the index notations $\a=+,-$ and $m=++,--$.}
\begin{align}
    \mathcal{L}_0 \longrightarrow \mathcal{L}_{\text{SUGRA}} &= \left( \int d^4 \theta \, E^{-1} \,
     \mathcal{A} ( \bbPhi, \nabla_{\pm} \bbPhi , \nabla_+ \nabla_- \bbPhi, \nabla_{\pm \pm} \bbPhi , \text{c.c.} ) \right)
      \nonumber \\
    &\quad 
    + \left(
    \int d^2 \theta \, \mathcal{E}^{-1} \, W ( \bbPhi ) 
    \right) 
    + \left( 
    \int d^2 \thetab \, \overline{\mathcal{E}}^{-1} \, 
    \Wb ( \bbPhib ) 
    \right)~ .
    \label{general_lag_coupled}
\end{align}
Here $\nabla_{\pm}$ is the derivative which is covariant with respect to the full local supergravity gauge group, 
$E^{-1}$ is the full superspace measure, $\mathcal{E}^{-1}$ is the chiral measure, and $\bbPhi$ is the covariantly 
chiral version of the chiral superfield $\Phi$---that is, $\nablab_{\pm} \bbPhi = 0$ whereas $\Db_{\pm} \Phi = 0$.

Expressions for these supercovariant derivatives and measures have been worked out in a series of papers
 \cite{Grisaru:1994dm,Grisaru:1995dr,Grisaru:1995kn,Grisaru:1995py,Gates:1995du} from which we will import the results
that we need for our analysis. 
To leading order in $H^m$, 
the linearized inverse superdeterminant of the supervielbein is
\begin{align}
    E^{-1} = 1 - \left[ \Db_+ , D_+ \right] H^{++} - \left[ \Db_- , D_- \right] H^{--}
\end{align}
while the chiral measure is given by
\begin{align}
    \mathcal{E}^{-1} &= e^{- 2 \sigma } \left( 1 \cdot e^{iH^m\overset{\leftarrow}{\pa_m}} \right)
    =1-2\s+i(\pa_m H^m)
    +\cdots
    ~ ,
\end{align}
where the ellipsis are terms of higher-order in $H^m$ and $\s$.
The covariantly chiral superfield $\bbPhi$ is related to the ordinary chiral superfield $\Phi$ by
\begin{align}
    \bbPhi = e^{i H^m\partial_m} \Phi = \Phi + i \left( H^{++} \partial_{++} + H^{--} \partial_{--} \right) \Phi + \mathcal{O} ( H^2 ) ~.
    \label{cov_chiral}
\end{align}
The spinor supercovariant derivatives $\nabla_{\pm}$ are
\begin{align}
    \nabla_\alpha = E_\alpha + \Omega_{\alpha} M + \Gamma_{\alpha} \Mb + \Sigma_{\alpha} N~ ,
    \label{supercovariant_deriv_grisaru}
\end{align}
where $M$ and $N$ are linear combinations of the Lorentz, $U(1)_V$, and $U(1)_A$ generators which act on spinors as
\bsubeq\begin{align}
    [ M , \psi_{\pm} ] &= \pm \frac{1}{2} \psi_{\pm}~,
    & [ M , \psib_{\pm} ] &= 0
    ~ ,\\
    [ \Mb , \psib_{\pm} ] &= \pm \frac{1}{2} \psib_{\pm}~,
    &  [ \Mb, \psi] &= 0~ , \\
    [ N , \psi_{\pm} ] &= - \frac{i}{2} \psi_{\pm}
    ~,& [N, \psib_{\pm}] &= + \frac{i}{2} \psib_{\pm}~ . 
\end{align}
\esubeq
The spinor inverse of the 
supervielbein $E_\alpha=E_\alpha{}^M\pa_M$,
and the structure group connections
$\Omega_{\alpha}$, $\Gamma_{\alpha}$, and $\Sigma_{\alpha}$ can be expressed to linear order in terms of the metric 
superfield $H^{\pm \pm}$ and an unconstrained complex
scalar compensator $S$. In the case of old-minimal supergravity,
the unconstrained superfield $S$ is related to the chiral compensator $\sigma$ by
\begin{align}
    S &= \sigma - \frac{i}{2} \partial_m H^m - \frac{1}{2} \left[ \Db_+ , D_+ \right] H^{++} - \frac{1}{2} \left[ \Db_- , D_- \right] H^{--}~ ,
    \label{S_sigma_contraint}
\end{align}
to linear order.
In the following analysis we will first obtain expressions for the supercovariant derivatives in terms of 
$S=S(H^m,\s)$, 
and use \eqref{S_sigma_contraint} 
to give them in terms of $H^m$ and $\sigma$. 

The spinorial inverse of the supervielbein is given at first order in the prepotentials by
\be  
E_\pm 
=
(1+\Sb)D_\pm 
+ i ( D_\pm H^m ) \partial_m 
- 2 \left( \Db_\mp D_\pm H^{\mp\mp} \right) D_\mp 
~,
\label{spin_viel}
\ee 
together with their complex conjugates.
Meanwhile, the connections $\Omega_\alpha$, $\Gamma_\alpha$, and $\Sigma_\alpha$ can be written to leading order as
\bsubeq
  \label{gammas_sigmas_omegas}
\bea
    \Gamma_\pm 
    &=&
    \pm2 D_\pm \left( S + \Sb \right) 
    \mp 2 D_\mp \Db_\mp D_\pm H^{\mp\mp} 
    ~,\\
    \Sigma_\pm &=& 
    - 2 i D_\pm \Sb + 2 i D_\mp \Db_\mp D_\pm H^{\mp\mp} 
    ~,
    \\
    \Omega_\pm &=& \mp 2 D_\mp \Db_\mp D_\pm H^{\mp\mp} 
    ~ .
\eea\esubeq
Using (\ref{supercovariant_deriv_grisaru}), the vielbeins (\ref{spin_viel}), and the expression (\ref{cov_chiral}) for $\bbPhi$, 
we find the supercovariant derivatives
\bsubeq    \label{linear_first_supercov_deriv}
\bea
    \nabla_\pm \bbPhi &=& 
    \left( 1 + \Sb \right) D_\pm \Phi 
    + 2 i ( D_\pm H^m ) \partial_m \Phi 
    + i H^m ( D_\pm \partial_m \Phi ) 
    - 2 \left( \Db_\mp D_\pm H^{\mp\mp} \right) D_\mp \Phi
    ~ , \\
     \nablab_\pm \bbPhib 
    &=&
    \left( 1 + S \right) \Db_\pm \Phib 
    - 2 i \left( \Db_\pm  H^m \right) \partial_m \Phib 
    - i H^m ( \Db_\pm \partial_m \Phib ) 
    - 2 \left( \Db_\pm D_\mp H^{\mp\mp} \right) \Db_\mp  \Phib 
    ~.~~~~~~~~~~~
\eea
\esubeq
To compute the second supercovariant derivatives
acting on $\bbPhi$ and $\bbPhib$, we must include the contributions from 
$\Omega_{\alpha}$, $\Gamma_{\alpha}$, $\Sigma_{\alpha}$, and their conjugates. One finds
\bsubeq
\begin{align}
    \nablab_+ \nabla_+ \bbPhi &= i ( 1 + S + \Sb ) \partial_{++} \Phi     
    - 2 ( \Db_+ \Db_- D_+ H^{--} ) D_- \Phi
+ 2 i ( \Db_+ D_+ H^m ) \partial_m \Phi  
    \non\\
    &\quad 
    - H^m \partial_{++} \partial_m \Phi 
    + 2 \left( \Db_+ ( S + \Sb ) + \Db_- D_- \Db_+ H^{--} \right) D_+ \Phi
    ~,  \\
    \nabla_+ \nabla_- \bbPhi &= 
    ( 1 + 2 \Sb )  D_+ D_- \Phi 
    + 2 i ( D_+ D_- H^m ) \partial_m \Phi 
    - 2 i ( D_- H^m )  D_+ \partial_m \Phi 
    \non\\
    &\quad
    + 2 i ( D_+ H^m )  D_- \partial_m \Phi
    + i H^m  D_+ D_- \partial_m \Phi  
    - 2 ( D_+ \Db_+ D_- H^{++} ) D_+ \Phi 
    \non\\
    &\quad
    + 2 \left( D_- \Db_- D_+ H^{--} \right) D_- \Phi 
    ~,  \\
    \nablab_- \nabla_- \bbPhi 
    &=
    i ( 1 + S + \Sb ) \partial_{--} \Phi 
    - 2 ( \Db_- \Db_+ D_- H^{++} ) D_+ \Phi
    + 2 i ( \Db_- D_- H^m ) \partial_m \Phi 
       \non\\
    &\quad 
    - H^m \partial_{--} \partial_m \Phi
    + 2 \left( \Db_- ( S + \Sb ) + \Db_+ D_+ \Db_- H^{++} \right) D_- \Phi
    ~,
\end{align}
\esubeq
together with their complex conjugates.
Armed with these expressions, we can linearize the supergravity couplings in (\ref{general_lag_coupled}). 
First let us consider the contribution from the D-term. We would like to extract the terms proportional to $H^{\pm \pm}$ and $\sigma$ in
\begin{align}
    \mathcal{L} =&
    ~
     \int d^4 \theta \, E^{-1} \, \mathcal{A} ( \bbPhi, \nabla_{\pm} \bbPhi , \nabla_+ \nabla_- \bbPhi, \nabla_{\pm \pm} \bbPhi , \text{c.c.} ) 
    ~, \nonumber \\
    &\sim 
    \int d^4 \theta \Big( H^{\alpha \alphad} [ D_{\alpha}, \Db_{\alphad} ] \mathcal{A} 
    + i \frac{\partial \mathcal{A}}{\partial \Phi} H^m \partial_m \Phi + \left( \nabla_\alpha \bbPhi - D_\alpha \Phi \right) 
     \frac{\partial \mathcal{A}}{\partial \nabla_{\alpha} \Phi}
      \nonumber \\
    &
    + \frac{\partial \mathcal{A}}{\partial \nabla_{+} \nabla_{-} \Phi} \left( \nabla_+ \nabla_- \bbPhi - D_+ D_- \Phi \right) 
    + \frac{\partial \mathcal{A}}{\partial \nabla_m \Phi} \left( \nabla_m \bbPhi - \partial_m \Phi \right) + \text{c.c.} \Big) 
    ~,
\end{align}
where $\nabla_{\pm \pm} = - i \left\{ \nabla_{\pm} , \nablab_{\pm} \right\}$. Doing so, we see that the currents which couple to 
$H^{\pm \pm}$ are
\begin{align}\hspace{-50pt}
    \mathcal{J}_{++} &=
    [D_+, \Db_+] \Bigg[\, 
    \frac{1}{2} \mathcal{A}
    - \frac{1}{2}   \frac{\partial \mathcal{A}}{\partial \nabla_{-} \Phi}  D_- \Phi
    - \frac{1}{2} \frac{\partial \mathcal{A}}{\partial \nabla_{+} \Phi} D_+ \Phi
    +  \frac{\partial \mathcal{A}}{\partial \nabla_{+} \nabla_{-} \Phi}D_+ D_- \Phi 
    +  \frac{\partial \mathcal{A}}{\partial \nabla_{++} \Phi}\partial_{++} \Phi 
    \nonumber \\
    &\hspace{60pt}   
    + 2 i \Db_- \left(\frac{\partial \mathcal{A}}{\partial \nabla_{--} \Phi}  D_- \Phi\right) 
    + 2 i  \Db_+ \left(\frac{\partial \mathcal{A}}{\partial \nabla_{++} \Phi}D_+ \Phi \right) 
    + \frac{\partial \mathcal{A}}{\partial \nabla_{--} \Phi}
    \partial_{--} \Phi \,\Bigg]
 \nonumber \\
    &\quad + i \Bigg[\,
    \frac{\partial \mathcal{A}}{\partial \Phi} \partial_{++} \Phi 
    + \frac{1}{2} \partial_{++} \left(\frac{\partial \mathcal{A}}{\partial \nabla_{-} \Phi} D_- \Phi  \right)
    - \partial_{++} \left(\frac{\partial \mathcal{A}}{\partial \nabla_+ \nabla_- \Phi} D_+ D_- \Phi \right) 
    -\frac{\partial \mathcal{A}}{\partial \nabla_{-} \Phi}  D_- \partial_{++} \Phi  \nonumber \\
    &\hspace{30pt} 
    - 2 D_- \left(\frac{\partial \mathcal{A}}{\partial \nabla_{-} \Phi} \partial_{++} \Phi\right)
    + \frac{1}{2} \partial_{++} \left(\frac{\partial \mathcal{A}}{\partial \nabla_{+} \Phi}D_+ \Phi  \right)  
    - \frac{\partial \mathcal{A}}{\partial \nabla_+ \Phi}D_+ \partial_{++} \Phi
    - 2 D_+ \left( \frac{\partial \mathcal{A}}{\partial \nabla_+ \Phi} \partial_{++} \Phi\right) 
    \nonumber \\
    &\hspace{30pt} 
    - 2 D_- D_+ \left(  \frac{\partial \mathcal{A}}{\partial \nabla_{+} \nabla_{-} \Phi}\partial_{++} \Phi \right) 
    + 2 D_- \left(  \frac{\partial \mathcal{A}}{\partial \nabla_{+} \nabla_{-} \Phi} D_+ \partial_{++} \Phi  \right)
    \nonumber \\
    &\hspace{30pt} 
    - 2 D_+ \left(  \frac{\partial \mathcal{A}}{\partial \nabla_{+} \nabla_{-} \Phi}D_- \partial_{++} \Phi \right)
    + \frac{\partial \mathcal{A}}{\partial \nabla_{+} \nabla_{-} \Phi} D_+ D_- \partial_{++} \Phi 
    + 2 i D_+ \Db_+ \left(\frac{\partial \mathcal{A}}{\partial \nabla_{++} \Phi} \partial_{++} \Phi  \right) 
    \nonumber \\
    &\hspace{30pt} 
    + 2 i D_- \Db_- \left(  \frac{\partial \mathcal{A}}{\partial \nabla_{--} \Phi}\partial_{++} \Phi \right) 
    + \frac{\partial \mathcal{A}}{\partial \nabla_{++} \Phi}\partial_{++}^2 \Phi 
    +\frac{\partial \mathcal{A}}{\partial \nabla_{--} \Phi}\partial_{--} \partial_{++} \Phi \,\Bigg]
    \nonumber \\
    &\quad 
    + 2 \Bigg[ \,- D_- \Db_+ \left( \frac{\partial \mathcal{A}}{\partial \nabla_{-} \Phi}D_+ \Phi \right) 
    - D_- \Db_+ D_+ \left(  \frac{\partial \mathcal{A}}{\partial \nabla_{+} \nabla_{-} \Phi}D_+ \Phi \right)
    \nonumber \\
    &\hspace{40pt} 
    + i D_- \Db_+ \Db_- \left(   \frac{\partial \mathcal{A}}{\partial \nabla_{--} \Phi}D_+ \Phi \right)  
    - i \Db_- D_+ \Db_+ \left(  \frac{\partial \mathcal{A}}{\partial \nabla_{--} \Phi}D_- \Phi \right) \Bigg] 
    \nonumber \\
    &\quad + \text{ c.c. } ~,
    \label{general_jpp}
\end{align}
and
\begin{align}\hspace{-50pt}
    \mathcal{J}_{--} &= [D_-, \Db_-] \Bigg[\,
    \frac{1}{2} \mathcal{A} 
    - \frac{1}{2}  \frac{\partial \mathcal{A}}{\partial \nabla_{-} \Phi}D_- \Phi
    - \frac{1}{2} \frac{\partial \mathcal{A}}{\partial \nabla_{+} \Phi} D_+ \Phi 
    +  \frac{\partial \mathcal{A}}{\partial \nabla_{+} \nabla_{-} \Phi} D_+ D_- \Phi
    + \frac{\partial \mathcal{A}}{\partial \nabla_{++} \Phi} \partial_{++} \Phi 
    \nonumber \\
    &\hspace{60pt}  
    + 2 i \Db_- \left(  \frac{\partial \mathcal{A}}{\partial \nabla_{--} \Phi}D_- \Phi  \right) 
    + 2 i  \Db_+ \left(   \frac{\partial \mathcal{A}}{\partial \nabla_{++} \Phi}D_+ \Phi \right) 
    + \frac{\partial \mathcal{A}}{\partial \nabla_{--} \Phi}\partial_{--} \Phi 
    \,\Bigg] \nonumber \\
    &\quad + i \Bigg[\,
     \frac{\partial \mathcal{A}}{\partial \Phi} \partial_{--} \Phi 
     + \frac{1}{2} \partial_{--} \left(   \frac{\partial \mathcal{A}}{\partial \nabla_{-} \Phi} D_- \Phi \right) 
     - \partial_{--} \left( \frac{\partial \mathcal{A}}{\partial \nabla_+ \nabla_- \Phi}D_+ D_- \Phi  \right)  
     -  \frac{\partial \mathcal{A}}{\partial \nabla_{-} \Phi} D_- \partial_{--} \Phi
     \nonumber \\
    &\hspace{30pt} 
- 2 D_- \left(  \frac{\partial \mathcal{A}}{\partial \nabla_{-} \Phi} \partial_{--} \Phi\right)
+ \frac{1}{2} \partial_{--} \left(   \frac{\partial \mathcal{A}}{\partial \nabla_{+} \Phi} D_+ \Phi \right)  
- \frac{\partial \mathcal{A}}{\partial \nabla_+ \Phi}D_+ \partial_{--} \Phi
- 2 D_+ \left(  \frac{\partial \mathcal{A}}{\partial \nabla_+ \Phi} \partial_{--} \Phi\right) 
\nonumber \\
    &\hspace{30pt} 
    - 2 D_- D_+ \left(  \frac{\partial \mathcal{A}}{\partial \nabla_{+} \nabla_{-} \Phi} \partial_{--} \Phi\right) 
    + 2 D_- \left(  \frac{\partial \mathcal{A}}{\partial \nabla_{+} \nabla_{-} \Phi}D_+ \partial_{--} \Phi  \right) 
    \nonumber \\
    &\hspace{30pt} 
    - 2 D_+ \left(\frac{\partial \mathcal{A}}{\partial \nabla_{+} \nabla_{-} \Phi}D_- \partial_{--} \Phi\right)
    +\frac{\partial \mathcal{A}}{\partial \nabla_{+} \nabla_{-} \Phi} D_+ D_- \partial_{--} \Phi 
    + 2 i D_+ \Db_+ \left(  \frac{\partial \mathcal{A}}{\partial \nabla_{++} \Phi}\partial_{--} \Phi \right)  
    \nonumber \\
    &\hspace{30pt}
    + 2 i D_- \Db_- \left(  \frac{\partial \mathcal{A}}{\partial \nabla_{--} \Phi} \partial_{--} \Phi\right) 
    + \frac{\partial \mathcal{A}}{\partial \nabla_{++} \Phi} \partial_{++} \partial_{--} \Phi 
    +  \frac{\partial \mathcal{A}}{\partial \nabla_{--} \Phi} \partial_{--}^2 \Phi\,  \Bigg] 
    \nonumber \\
    &\quad + 2 \Bigg[\,
     -D_+ \Db_- \left(   \frac{\partial \mathcal{A}}{\partial \nabla_{+} \Phi} D_- \Phi\right) 
     + D_+ \Db_- D_- \left(  \frac{\partial \mathcal{A}}{\partial \nabla_{+} \nabla_{-} \Phi} D_- \Phi  \right) 
     \nonumber \\
    &\hspace{40pt} 
    + i D_+ \Db_- \Db_+ \left(  \frac{\partial \mathcal{A}}{\partial \nabla_{++} \Phi}  D_- \Phi\right) 
    - i \Db_+ D_- \Db_- \left(  \frac{\partial \mathcal{A}}{\partial \nabla_{++} \Phi}D_+ \Phi  \right) \Bigg] \nonumber \\
    &\quad + \text{ c.c. }~ ,
    \label{general_Jmm}
\end{align}
where $+\text{c.c.}$ means to add the complex conjugates of \emph{all} preceding terms (including the real quantity 
$\frac{1}{2} [ D_{\pm} , \Db_{\pm} ] \mathcal{A}$ for which the complex conjugate merely removes the factor of $\frac{1}{2}$).

The field $\mathcal{V}$ which appears in our deformation (\ref{FZTTbar}) receives two contributions, one from the D-term coupling 
which depends only on $\mathcal{A}$, and one from the F-term coupling which depends only on the superpotential $W$. 
Adding them, we find
\bea
    \mathcal{V}
    &=&
    \Db_+ \Db_- \Bigg[\,
    -\frac{\partial \mathcal{A}}{\partial \nabla_{\alpha} \Phi}  D_{\alpha} \Phi 
    + 2 \frac{\partial \mathcal{A}}{\partial \nabla_+ \nabla_- \Phi}D_+ D_- \Phi 
    + \frac{\partial \mathcal{A}}{\partial \nabla_{m} \Phi} \partial_{m} \Phi 
    +  \frac{\partial \mathcal{A}}{\partial \nabla_{m} \Phib}\partial_{m} \Phib
    \non\\
    &&~~~~~~~~~~
    + 2 i \Db_{+} \left(\frac{\partial \mathcal{A}}{\partial \nabla_{++} \Phi}  D_+ \Phi \right) 
    + 2 i D_{+} \left(  \frac{\partial \mathcal{A}}{\partial \nabla_{++} \Phib} \Db_{+} \Phib\right)
    \non\\
    &&~~~~~~~~~~
    + 2 i \Db_{-} \left( \frac{\partial \mathcal{A}}{\partial \nabla_{--} \Phi} D_- \Phi \right) 
    + 2 i D_{-} \left( \frac{\partial \mathcal{A}}{\partial \nabla_{--} \Phib}\Db_{-} \Phib  \right)
    \Bigg]
    \non\\
    &&
    + 2 W ( \Phi )~ .
\eea

\section{Simplifying the deformation on-shell}\label{appendix:on-shell}

In this Appendix, we prove the claim that one can drop all terms which involve products of 
$(D_+ D_- \Phi)$ or 
$(\Db_+ \Db_- \Phib)$ and the four-fermion term $|D\Phi|^4 = D_+ \Phi \Db_+ \Phib D_- \Phi \Db_- \Phib$ 
when the equations of motion are satisfied.

To see this for the models we consider,  
it suffices to consider a superspace Lagrangian of
the form
\begin{align}
    \mathcal{L} &= \int d^4 \theta \, \mathcal{A}
     \left( \Phi, D_{\pm} \Phi, D_+ D_- \Phi, \partial_{\pm \pm} \Phi , \text{c.c.} \right) \nonumber \\
    &= \int d^4 \theta \, \left( K ( \Phi , \Phib ) + f ( x, \xb, y ) |D\Phi|^4 \right) ~,
\end{align}
which has the superspace equation of motion
\begin{align}
    \Db_+ \Db_- K_{\Phi} = \Db_+ \Db_- \Bigg\{&
     D_{\alpha} \left[ \frac{\partial ( f |D\Phi|^4 ) }{\partial D_{\alpha} \Phi} \right]
     - D_+ D_- \left[ \frac{\partial ( f |D\Phi|^4 )}{\partial D_+ D_- \Phi} \right] 
     \cr &
     - \partial_{m} \left[\frac{\partial ( f |D\Phi|^4 )}{\partial ( \partial_{m} \Phi ) } \right] \Bigg\}
    \label{superspace_eom}
\end{align}
for $\Phi$, and the conjugate equation of motion for $\Phib$. 
If we multiply (\ref{superspace_eom}) on both sides by the four-fermion term 
$|D\Phi|^4 = D_+ \Phi \Db_+ \Phib D_- \Phi \Db_- \Phib$ then any term containing $(D_\pm\Phi)$ and
$(\Db_\pm\overline{\Phi})$
fermions in (\ref{superspace_eom}) will vanish by nilpotency. On the left, the only surviving term is 
$K_{\Phi \Phib} \Db_+ \Db_- \Phib$, while on the right we get contributions from the first and second terms:
\begin{align}
     K_{\Phi \Phib} \left(\Db_+ \Db_- \Phib \right) |D\Phi|^4 = 
     \left( \Db_+ \Db_- \Phib \right) |D\Phi|^4
     \Bigg\{&  \lambda  \Db_+ \Db_- \left[ \frac{\partial f}{\partial y}  ( \partial_{--} \Phib ) ( \partial_{++} \Phib ) \right]
   \cr & 
   - \left( \frac{x + \xb}{\lambda} \right) f\Bigg\}  ~.
\end{align}
On collecting terms, the previous equation turns into
\begin{align}
   \left( \Db_+ \Db_- \Phib \right) |D\Phi|^4 \left\{
     K_{\Phi \Phib} + \left( \frac{x + \xb}{\lambda} \right) f 
     - \lambda  \Db_+ \Db_- \left[ \frac{\partial f}{\partial y}  ( \partial_{--} \Phib ) ( \partial_{++} \Phib ) \right] \right\}  
     = 0 ~. 
     \label{C4}
\end{align}
The parenthesis multiplying 
$( \Db_+ \Db_- \Phib )|D\Phi|^4$ 
in the previous expression
does not vanish in general,
at least for 
$\lambda$ small enough.
Then for \eqref{C4} to be satisfied,  
the equation
\begin{align}
    \left( \Db_+ \Db_- \Phib \right) |D\Phi|^4 = 0  
\end{align}
has to hold
when the equations of motion are satisfied. This justifies our claim in section \ref{subsec:kahler}
 that we may drop all terms involving the product $y |D\Phi|^4$ in the deformation, assuming we restrict to on-shell configurations.

\bibliographystyle{utphys}
\bibliography{master}

\end{document}